\documentclass[a4paper,11pt]{article}
\pdfoutput=1 

\usepackage{jcappub,ulem} 
\usepackage{float}
\usepackage[T1]{fontenc} 
\usepackage[utf8]{inputenc}
\usepackage{orcidlink}
\usepackage{graphicx}
\usepackage{natbib}
\usepackage{subfigure}
\title{ WIMP Dark Matter Searches in Reticulum II Using MeerKAT}

\author[1,*]{Shibre Semane\orcidlink{0009-0000-1838-4469},\note[*]{Corresponding author}}
\author[1]{Geoff Beck \orcidlink{0000-0003-4916-4914},}
\author[1,2]{Sphesihle Makhathini \orcidlink{0000-0001-9565-9622},}
\author[3,4]{Marco Regis \orcidlink{0000-0003-0399-0284}}
\author[5,6,7]{and Gianni Bernardi \orcidlink{0000-0002-0916-7443}}

\affiliation[1]{Wits Centre for Astrophysics (Wits CfA) and School of Physics, University of the Witwatersrand, Johannesburg, Wits 2050, South Africa}
\affiliation[2]{INAF – Osservatorio Astronomico di Cagliari, via della Scienza 5, 09047, Selargius (CA), Italy}
\affiliation[3]{Dipartimento di Fisica, Università di Torino,
via P. Giuria 1, I–10125 Torino, Italy}
\affiliation[4]{Istituto Nazionale di Fisica Nucleare, Sezione di Torino, via P. Giuria 1, I–10125 Torino, Italy}
\affiliation[5]{INAF — Istituto di Radioastronomia, Via Gobetti 101, 40129, Bologna, Italy}
\affiliation[6]{Centre for Radio Astronomy Techniques and Technologies (RATT), Department of Physics and Electronics, Rhodes University, Makhanda 6140, South Africa}
\affiliation[7]{South African Radio Astronomy Observatory, Cape Town 7700, South Africa}

\emailAdd{shibre.semane@gmail.com}
\emailAdd{geoffrey.beck@wits.ac.za}
\emailAdd{sphemakh@gmail.com}
\emailAdd{marco.regis@unito.it}
\emailAdd{giannibernardi75@gmail.com}

\abstract{In the last decade radio astronomy has emerged as a powerful technique for detecting signatures of Weakly Interacting Massive Particles (WIMPs). Dwarf spheroidal galaxies (dSphs) are particularly promising targets for these searches due to their substantial dark matter (DM) dominance and minimal baryonic background emission.
In this study, we utilize the exceptional sensitivity of the MeerKAT radio telescope to search for synchrotron emission from WIMP annihilation/decay in the nearby \textsc{Reticulum II} dSph. Through rigorous data reduction and self-calibration, we establish constraints on WIMP properties that improve upon 
previous radio studies, demonstrating the potential of MeerKAT and next-generation radio telescopes in exploring increasing swathes of the WIMP parameter space. 

}

\begin{document}
\flushbottom
\maketitle

\section{Introduction}

Dark matter (DM) indirect detection studies continue to evolve alongside the instruments we use to carry them out. As modern telescopes, such as the MeerKAT ~\cite{jonas2016meerkat}, and ASKAP~\cite{hotan2021australian}, both precursors to the Square Kilometre Array (SKA), as well as the Low Frequency Array (LOFAR)~\cite{rottgering2003lofar} radio telescopes, continue to push the sensitivity boundary, we are able to probe DM parameter spaces more deeply. 

Local Group dwarf spheroidal galaxies (dSphs) have long served as testbeds for several key astrophysical questions — from understanding structure formation and galaxy evolution \cite{white1978core}, to testing the DM hypothesis due to their extremely high DM densities~\cite{mateo1998dwarf,strigari2018dark}, and addressing the persistent gap between the number of satellite galaxies predicted by $\Lambda$CDM simulations and those actually observed \cite{klypin1999missing,moore1999dark}.\\
Since their discovery \cite{mateo1998dwarf, tolstoy2009star}, dSphs have proven to be fertile systems, studied across optical, X-ray, and radio wavelengths. With the advent of large-area surveys, like SDSS \cite{york2000sloan} and DES \cite{des2016overview}, dozens of new dSphs have been found and characterized, broadening the scope for testing DM models.
For DM studies, dSphs stand out due to their proximity and their extremely high mass-to-light ratios, often exceeding $ > 100\,M_\odot/L_\odot$, and in some cases reaching values near 500 \cite{strigari2008common, simon2019faintest,battaglia2022stellar}. These ratios are derived from a combination of stellar velocity dispersion measurements and Jeans analysis. They consistently point toward the presence of a substantial, unseen mass component, with no explanation other than DM.

Among the leading DM candidates, the Weakly Interacting Massive Particles (WIMPs) could annihilate/decay in the halos of dSphs, 
producing a population of secondary $e^\pm$, among other high-energy particles such as photons and neutrinos. In the presence of magnetic fields, these $e^\pm$ gyrate and emit synchrotron radiation, which appears as diffuse radio emission.
The transport and cooling of these $e^\pm$ is modelled using the diffusion-loss equation described in section 2, from which the synchrotron emission is computed.

Radio frequency searches for synchrotron emission from WIMPs in dSphs can probe mass ranges over $\sim$GeV-TeV, with the SKA forecasts extending sensitivity to TeV-scale and even trans-TeV WIMPs \cite{ chen2021sensitivity,kar2019constraints}.
Current constraints in these mass ranges mainly come from both gamma-ray and radio observations. On the gamma side, Fermi-LAT analyses of dSphs remain the benchmark limits \cite{mcdaniel2024legacy,abdollahi2026combined}.
On the radio side, existing dSph searches with GBT, GMRT, MWA, LOFAR, and ATCA have put competitive bounds especially for sub - 100 GeV WIMP masses \cite{spekkens2013deep, cook2020searching,kar2019constraints, basu2021stringent,gajovic2023weakly,regis2014local}.
Although constraints can be derived for all elementary particles as final states of annihilation, they are often reported for $b\bar{b}$, $\tau^+\tau^-$ and $\mu^+\mu^-$. The reason is that the spectra from different colored states and heavy gauge bosons produce similar $e^+e^-$ spectra, since related to the production and decay of charged pions, and thus $b\bar{b}$ is a representative example. $\tau^+\tau^-$ is a leptonic channel but with a semi-hadronic decay into pions, while $\mu^+\mu^-$ is purely leptonic, proceeding with a direct production of $e^+e^-$ through decays.
%

In this paper, we present upper limits on the WIMP annihilation cross-section/decay rate on three annihilation/decay channels ($b\bar{b}$, $\tau^+\tau^-$, and $\mu^+\mu^-$), based on MeerKAT observations of \textsc{Reticulum II} (\textsc{Ret II}). This dSph is an ultra-faint dwarf galaxy~\cite{bechtol2015eight} with a mass-to-light ratio approaching 500 \cite{koposov2015beasts, simon2015stellar}. \textsc{Ret II} lies at a distance of about 30 kpc, and has a compact angular extent — making it one of the most promising known dSphs for DM studies.
Despite being highly DM-dominated \cite{regis2017dark,koposov2015kinematics,bonnivard2015dark,simon2015stellar}, Ret II's magnetic field is poorly constrained and is a key source of uncertainty. For this reason, we compute expected emissions under optimistic and conservative self-confinement scenarios.

This work is a next-generation extension of the earlier Ret II ATCA's \cite{regis2017dark} work.
Compared to earlier radio dSph observations, the dedicated ATCA Ret II observation achieved competitive limits, excluding annihilation cross-sections around and below the thermal relic value depending on channel. 
The ATCA work demonstrated how single pointing, good source characterization and subtraction translate into tighter limits. 

We observed \textsc{Ret II} using MeerKAT, a precursor to SKA, for 10 hours in the UHF band (544–1088 MHz). 
The goal is to place upper limits on possible faint synchrotron emission arising from DM annihilation/decay. 
The data were calibrated, imaged, and self-calibrated to produce an optimal final map. Compact bright sources were then subtracted to create clean residuals. These residuals were analyzed for a potential DM signal by comparing them with the predicted synchrotron flux from WIMP models.
No significant excess emission was detected above the RMS level, allowing us to place upper limits on WIMP parameters. 
An upper limit that significantly improves upon previous radio searches~\cite{spekkens2013deep, natarajan2013bounds, basu2021stringent,kar2019constraints, gajovic2023weakly,regis2014local,vollmann2020radio}. 
Compared to ATCA, this work offers better sensitivity, coverage of the lower frequency band, more refined source subtraction and residual analysis, yielding significantly better limits (under the same model assumptions), showing the capability of MeerKAT, and paving the way to the SKA. 
We note that the strength of the resulting constraints is primarily determined by the achieved sensitivity, which, in this study, is on average 7.86 $\mu$Jy/beam.

The paper is structured as follows. Section~\ref{section:modeling} describes the model for synchrotron emission from DM. Section~\ref{section:datardx} details the MeerKAT observations and the data reduction process. The discussion of the results and the derived constraints are presented in Section~\ref{section:discres}. Section~\ref{section:conc} provides the conclusions.

\section{Emission Modeling}
\label{section:modeling}
There are multiple ingredients that go into predicting synchrotron emissions associated with DM annihilation or decay. Here we will outline our strategy for dealing with each aspect in turn.
\subsection{The Halo Profile}
The DM distribution within \textsc{Ret II} can be characterized using the Einasto profile \cite{einasto1974dynamic}:
\begin{equation}
\rho(r) = \rho_{-2} \exp\left\{-\frac{2}{\alpha} \left[\left(\frac{r}{r_{-2}}\right)^\alpha - 1\right]\right\},
\end{equation}
where $\alpha = 0.4$ is the shape parameter controlling how rapidly the slope changes with radius, $\rho_{-2} = 7.0 \times 10^{7}\ \mathrm{M_\odot\,kpc^{-3}}$
 is the characteristic density, and  $r_{-2} = 0.2$ kpc, the characteristic scale radius~\cite{bonnivard2015dark,regis2017dark}. It should be noted that using the median Navarro-Frenk-White profile from \cite{Pace_2018} instead does not substantially alter our results.
\subsection{Electron Production and Propagation}
The indirect detection of DM relies on a mulit-messenger approach, where different annihilation/decay by-products are observed via the corresponding instruments, such as prompt gamma rays in gamma-ray telescopes~\cite{mcdaniel2024legacy}, charged cosmic rays in cosmic-ray detectors~\cite{conrad2017indirect}, and neutrinos in neutrino telescopes~\cite{guepin2021indirect}. 
In the context of radio observations, however, the primary interest lies in the injection of high-energy electron and positron pairs ($e^\pm$).
In dSph halos, these $e^\pm$ are produced via WIMP annihilation/decay. 
%
For annihilation, the production rate scales as $\rho_\chi^2$, while for decay it scales as $\rho_\chi$~\cite{beck2019radio}. 
The resulting source functions are: 
\begin{equation}
   Q_e(r, E) = \frac{1}{2}\langle\sigma v\rangle\sum_f\frac{\mathrm{d}N^f_e}{\mathrm{d}E}\left(\frac{\rho_\chi(r)}{M_\chi}\right)^2
   \label{eq:source1}
\end{equation} and 
\begin{equation}
   Q_e(r, E) = \Gamma\sum_f\frac{\mathrm{d}N^f_e}{\mathrm{d}E}\left(\frac{\rho_\chi(r)}{M_\chi}\right)\;,
   \label{eq:source2}
\end{equation}
where $r$ is the distance from the center of Ret II, $E$ is the energy of the injected $e^\pm$, $\langle \sigma v \rangle$ is the velocity-averaged annihilation cross-section, $\Gamma$ is the decay rate, $M_{\chi}$ is WIMP mass, $\rho_{\chi}(r)$ is the DM density, given by the function discussed in the previous Section, and $\frac{dN^f_e}{dE}$ is the energy spectrum of $e^\pm$ originated from DM annihilating/decaying into the final state $f$. 
In this work, we focus on three benchmark annihilation/decay channels; $b\bar{b}$, $\tau^+\tau^-$, and $\mu^+\mu^-$. These are among the most explored final states in the literature, making them convenient for comparison. They are widely favored because they successfully span the spectrum from soft hadronic ($b\bar{b}$) to hard leptonic ($\tau^+\tau^-$ and $\mu^+\mu^-$) spectra and they're demonstrably constrainable via radio synchrotron searches \cite{kar2019constraints,regis2014local,basu2021stringent,regis2017dark}.
\subsection{Diffusion and Energy Losses} 

The propagation of the $e^\pm$ injected by WIMP annihilation/decay is described by the diffusion-loss equation~\cite{beck2023galaxy, colafrancesco2007detecting, mcdaniel2017multiwavelength}:
\begin{equation}
\frac{\partial}{\partial t} \psi_e = \nabla \left( D(E, r)\nabla\psi_e \right) + \frac{\partial}{\partial E} \left( b(E, r)\psi_e \right) + Q_e(E, r)\;,
\label{eq:diffusion}
\end{equation}
where $\psi_e$ is the $e^\pm$ equilibrium spectrum, we model the diffusion coefficient as $D(E, r) = D_0 \left( \frac{E}{1\ \mathrm{GeV}} \right)^{\delta} \left( \frac{B(r)}{B(0)} \right)^{-\delta}$ with $B$ being the magnetic field (and $\delta = 0.3$), $Q_e(E, r)$ is the source function in Eqs.~\eqref{eq:source1} and \eqref{eq:source2}, and $b(E, r)$ is the function describing energy losses, given by~\cite{sarkis2024darkmatters}: 
	\begin{equation}
	\begin{aligned}
		b(E, r) &= b_{\mathrm{IC}} \left( \frac{E}{1~\mathrm{GeV}} \right)^{2}
		+ b_{\mathrm{sync}} \left( \frac{E}{1~\mathrm{GeV}} \right)^{2}
		\left( \frac{B(r)}{1~\mu\mathrm{G}} \right)^{2} \\[4pt]
		&\quad + b_{\mathrm{Coul}}\!\left( \frac{n_e(r)}{1~\text{cm}^{-3}} \right)
		\!\left[1+\frac{\ln((E/\text{GeV})(1/n_e(r)))}{75}\right]
		+ b_{\mathrm{brem}}\, n_e(r)\, E.
	\end{aligned}
    \label{eq:bloss}
\end{equation}
The coefficients 
$b_{\mathrm{IC}}$, $b_{\mathrm{sync}}$, $b_{\mathrm{Coul}}$ and, $b_{\mathrm{brem}}$ correspond to the inverse Compton, synchrotron, Coulomb, and bremsstrahlung losses, respectively.
Given the low gas density in dSphs, we can neglect $b_{\mathrm{Coul}}$ and $b_{\mathrm{brem}}$.
The dominant loss is given by inverse Compton on CMB. For the magnetic field considered here ($\lesssim1\,\mu$G), the synchrotron radiation is just a small correction in Eq.~\ref{eq:bloss}, despite being the critical mechanism for radio observations and the primary focus of this study.\\
In our main analysis, we used the \texttt{DarkMatters}~\cite{sarkis2024darkmatters}~\footnote{\url{https://codeberg.org/Hyperthetical/DarkMatters}} package
to solve the diffusion-loss equation, calculate synchrotron emissivity, and generate surface brightness profiles for direct comparison with observed data.\\ 
\texttt{DarkMatters} implements an Operator-Splitting (OS) technique to solve the diffusion-loss equation.  
The OS method splits the full PDE in Eq.~(\ref{eq:diffusion}) into separate energy and spatial operators. These are applied in succession to generate a full time-step for the solution. We refer the reader to \cite{sarkis2024darkmatters} for a detailed explanation of the scheme.
\subsection{Synchrotron Emission}
Having obtained $\psi_e$, we calculate the synchrotron emissivity, i.e., the power emitted per unit volume per frequency:
\begin{equation}
j_{\text{synch}}(\nu, r) = \int_{m_e}^{M_\chi} dE \, \psi_e \left(E,r\right)\, P_{\text{synch}}(\nu, E, B),
\end{equation}
where $P_{\text{synch}}(\nu, E, B)$ is the synchrotron power at frequency $\nu$ emitted by an $e^\pm$ with energy $E$ in a magnetic field $B$.
The surface brightness is obtained by integrating the emissivity along the line of sight:

\begin{equation}
I_{\text{synch}}(\nu, R) = \int_{-\infty}^{\infty} dl \, j_{\text{synch}}(\nu, \sqrt{R^2 + l^2}),
\end{equation}
with $R$ being the projected radius and $l$ is the line-of-sight coordinate.
\subsection{Magnetic Field Assumptions}

The interpretation of dark-matter-induced synchrotron signals is highly sensitive to assumptions about the magnetic field, both for the total strength of the field, setting the synchrotron power, and for its turbulent properties, setting the diffusion of $e^\pm$. The magnetic field affects the size and the morphology of the predicted signal and thus the constraints derived on DM properties. However, the magnetic field strength in dSphs is poorly constrained, introducing significant uncertainty~\cite{regis2015local,regis2017dark,vollmann2020radio,gajovic2023weakly,kar2019constraints}.
Beyond the uncertainties in the magnetic field strength, another factor to consider is the spatial extent of the diffusion zone. For a dSph like \textsc{Ret II}, whose overall extent is very small, if the diffusion zone is comparable to the size of the object (i.e., hundreds of pc), the diffusion length of GeV electrons and positrons (which is a few kpc) is much longer than the confinement size, and thus a significant fraction of the power is carried away from the source. Thus, the larger the diffusion zone, the larger is the total emission power, and typically the stronger are the bounds.
In the following, we consider an optimistic and a conservative scenario in order to bracket such uncertainty in the magnetic field strength.
\subsubsection{A Spatially Uniform Magnetic Field}

As for the optimistic scenario, we assume a uniform magnetic field of strength $1~\mu$G~\cite{regis2017dark}, which in turn implies a uniform diffusion coefficient, and we set $D_0=3.0 \times 10^{28} \ \mathrm{cm}^{2} \ \mathrm{s}^{-1}$.
This can be seen as the assumptions that i) there is a large-scale magnetic field around the Milky-Way extending to the position of \textsc{Ret II} at the $\mu$G level, and ii) the level of turbulence in \textsc{Ret II} is similar to the one of the Milky-way (so to consider the same diffusion coefficient).
The halo component of the Milky-Way coherent magnetic field has a strength of a few $\mu$G and it is expected to extend several kpc away from the center \cite{Unger:2023lob}. On the other hand, \textsc{Ret II} is located at low declination, away from the disk, where the magnetic field strength is reduced. The assumption we make in the optimistic scenario is that this reduction is not dramatic.
The generation of turbulence is typically understood in connection to star formation, and thus the level of turbulence in \textsc{Ret II} is likely to be smaller than in the Milky-Way. We do not have a direct estimate of the magnetic turbulence based on observations (from the variance of rotation measure or other probes), which motivates us to take the diffusion coefficient of the Milky-Way.
This model yields the synchrotron emission in the left panel of Fig.~\ref{fig:self-uniform}.  
The upper limits derived from such synchrotron emission are shown in Fig.~\ref{fig:ULs_comp}.
Assuming uniform magnetic field and diffusion coefficient clearly maximises synchrotron emission by maintaining constant both the magnetic strength and the residence time of $e^\pm$. 
\label{subsec:uniform}
\subsubsection{Self-Confinement of Electrons and Positrons}

The injection of $e^\pm$ from DM annihilation generates irregularities in the magnetic field, giving rise to magnetic turbulence. In galaxies with significant star formation, this mechanism is highly subdominant with respect to the turbulence given by, e.g., supernova explosions.
On the other hand, \textsc{Ret II} is an ultra-faint dSph, and it is not unreasonable to consider stellar-driven turbulence to be absent.
In this conservative scenario, we thus consider only DM self-generated turbulence.
Ref.~\cite{regis2023self} showed that this mechanism leads to a non-negligible residence time of the $e^\pm$ in dSphs. 
Moreover, the dependence of this effect and of the synchrotron power on the strength of the coherent magnetic field is nearly opposite, thus canceling each other and making the prediction nearly independent from $B_0$, except for low DM masses~\cite{regis2023self}. We will come back to this point when discussing the results.
For the sake of definiteness, we assumed $B_0=0.1\,\mu$G.
This model sets a minimum "irreducible" synchrotron signal from WIMP DM and will be taken as our conservative scenario.
For details about the numerical solution of the transport equation, see ~\cite{regis2023self}.
\label{subsec:self-conf}
\raggedbottom
\subsubsection{Magnetic Field from the stellar-driven dynamo}

\begin{figure}[H]
        \centering
        \begin{minipage}{\textwidth}
        \centering
        \includegraphics[width=0.48\textwidth]{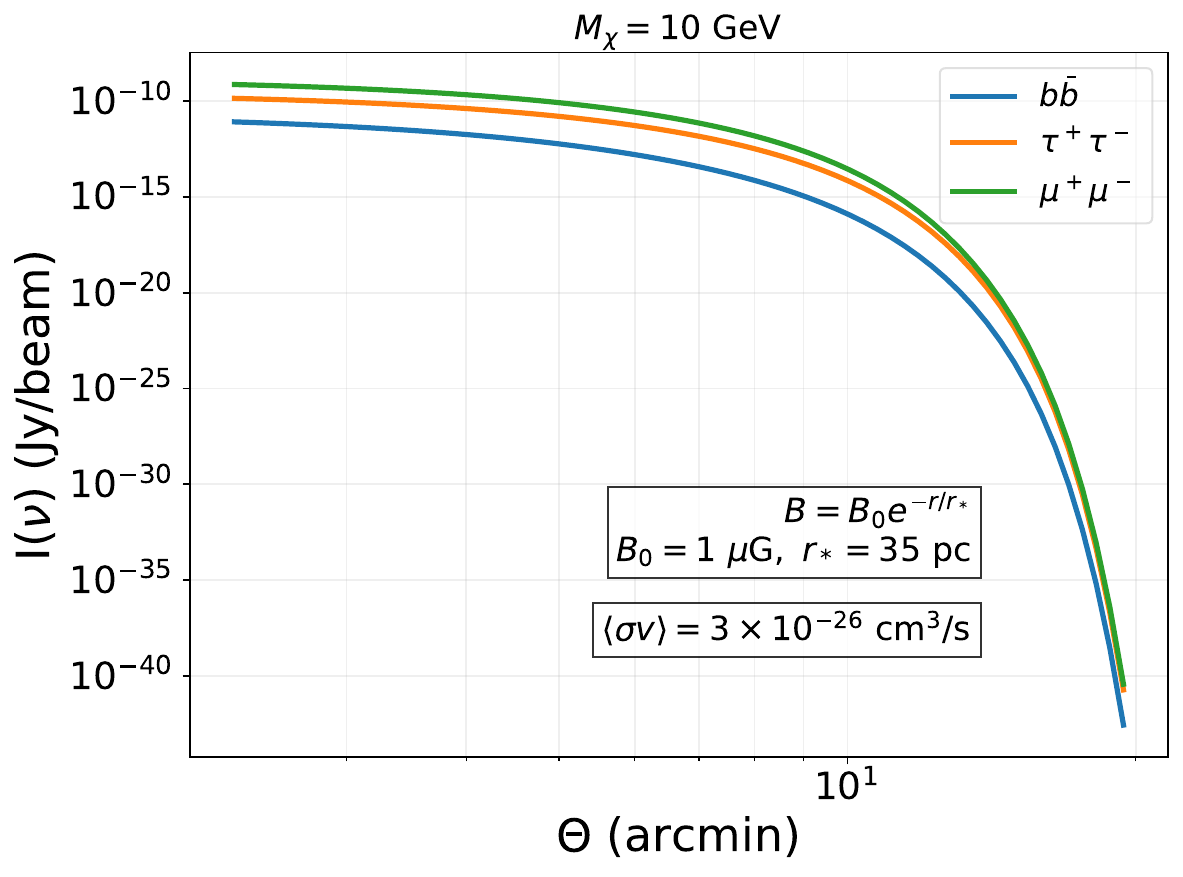}
        \includegraphics[width=0.48\textwidth]{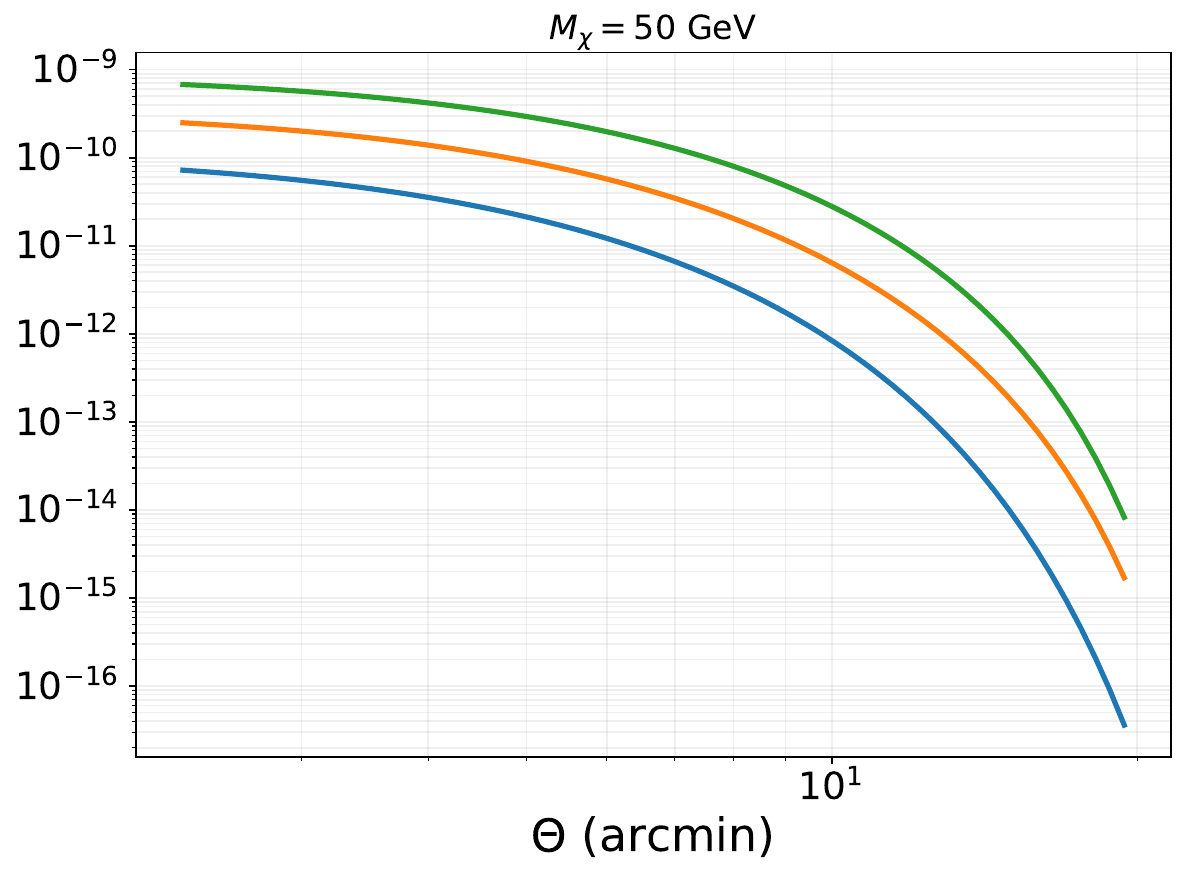}
        \includegraphics[width=0.48\textwidth]{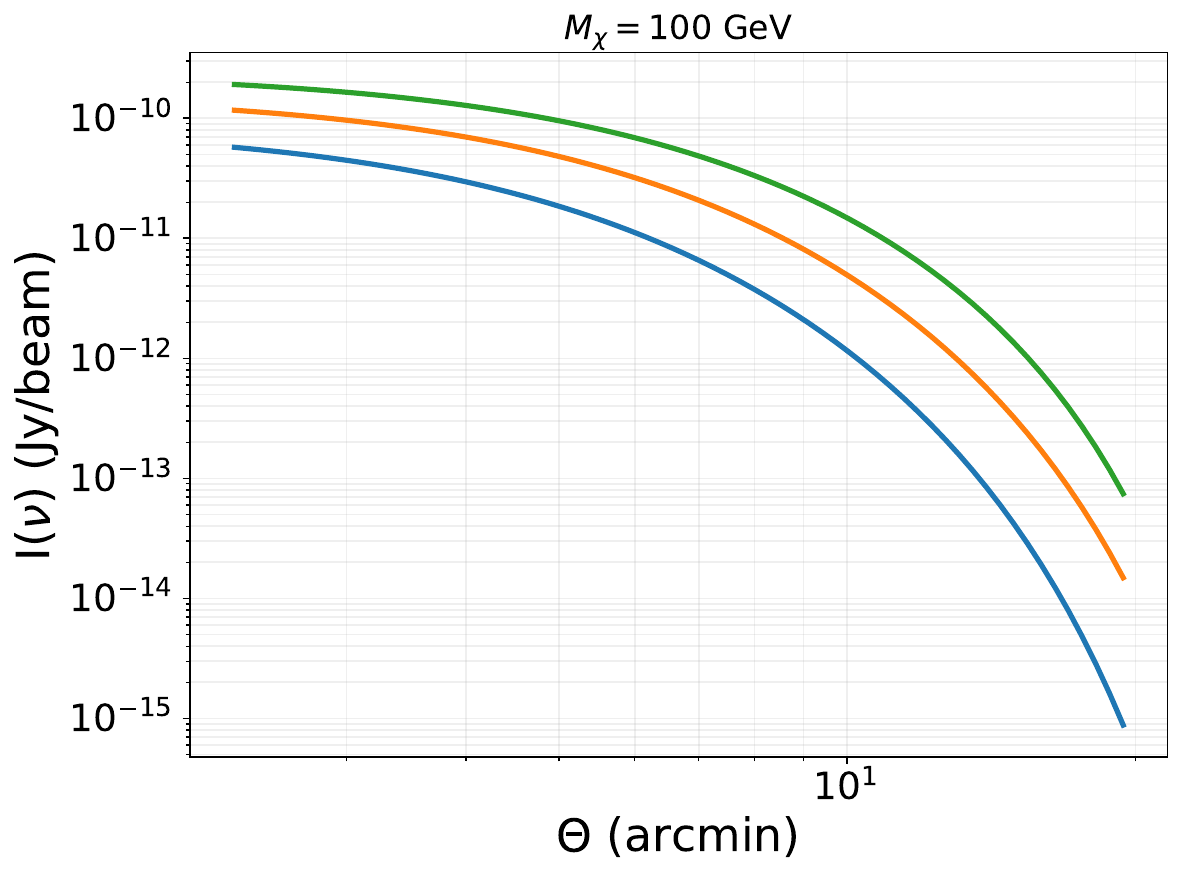}
        \includegraphics[width=0.48\textwidth]{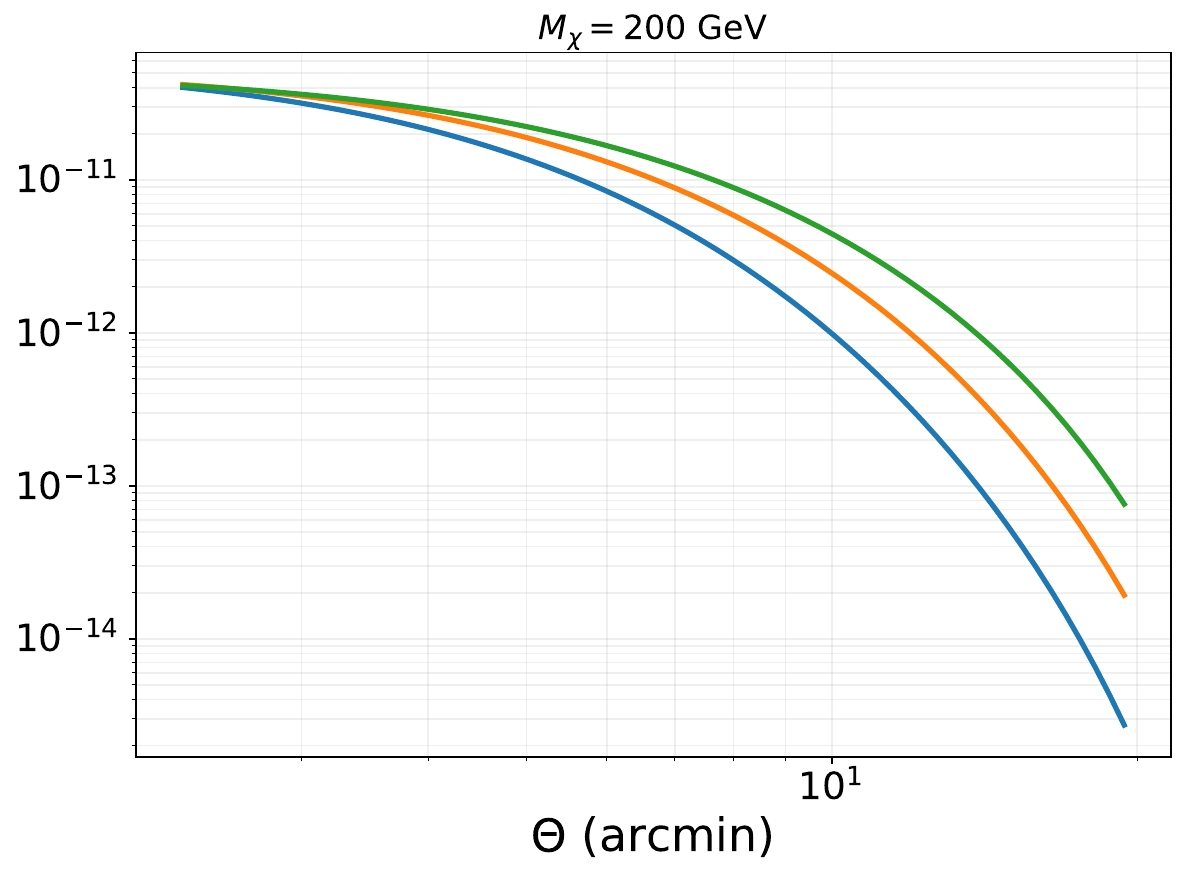}
    \end{minipage}
\caption{Radial brightness profiles computed for a few WIMP masses annihilating into $b\bar{b}$, $\tau^+\tau^-$, and $\mu^+\mu^-$ assuming an exponentially decaying magnetic field.}
\label{fig:expB}
\end{figure}

As already mentioned, the generation of magnetic fields in galaxies is typically understood in terms of dynamo mechanisms, sustained by energy from star formation-related processes~\cite{brandenburg2023galactic,mreider}.
In this case, it might be natural to model the magnetic spatial profile as $B(r) = B_0\,e^{-r/r_*}$, with $r_* = 35$ pc, given by the half-light radius of the \textsc{Ret II} stellar population. 
This spatial profile is adopted as a physically motivated assumption, drawing on the empirical trend observed in galaxies where magnetic field strength broadly traces stellar distribution. 
The same profile is then inherited by the diffusion coefficient. Given the small size of the \textsc{Ret II} galaxy, the strength is expected to be $B_0 <\mu\mathrm{G}$~\cite{regis2015local}.
The stellar population of \textsc{Ret II} is old, and an important question is whether a magnetic field generated at early times can be sustained until the recent epoch.
We set  $B_0 = 1\,\mu\mathrm{G}$ and plot the synchrotron emission obtained in this scenario in Fig.~\ref{fig:expB}.
Note that there are two effects depleting the signal: i) the confinement region is small (with size given by $r_*$) compared to the diffusion length, and a significant fraction of $e^\pm$ can escape, carry power away from the dwarf, and ii) the synchrotron emission volume (i.e., where $B$ is sizeable) is small (again, with size given by $r_*$).
The magnetic field from the stellar-driven dynamo would add to the DM self-generated magnetic field described above. Since the expected emission shown in Fig.~\ref{fig:expB} is lower than in the self-confinement case, and more uncertain, we will disregard the contribution. 

\begin{figure}[H]
    \centering
    \begin{minipage}{\textwidth}
        \centering
        \includegraphics[width=0.85\textwidth]{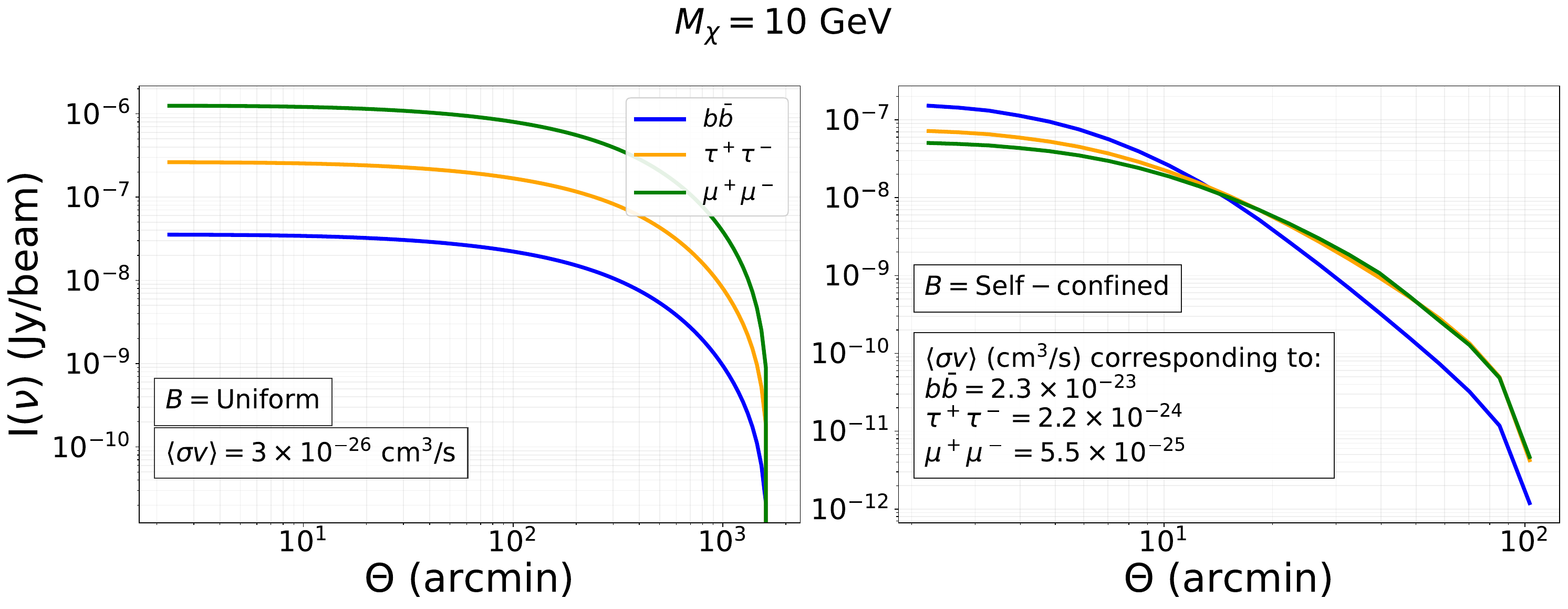}\\[0.4cm]
        \includegraphics[width=0.85\textwidth]{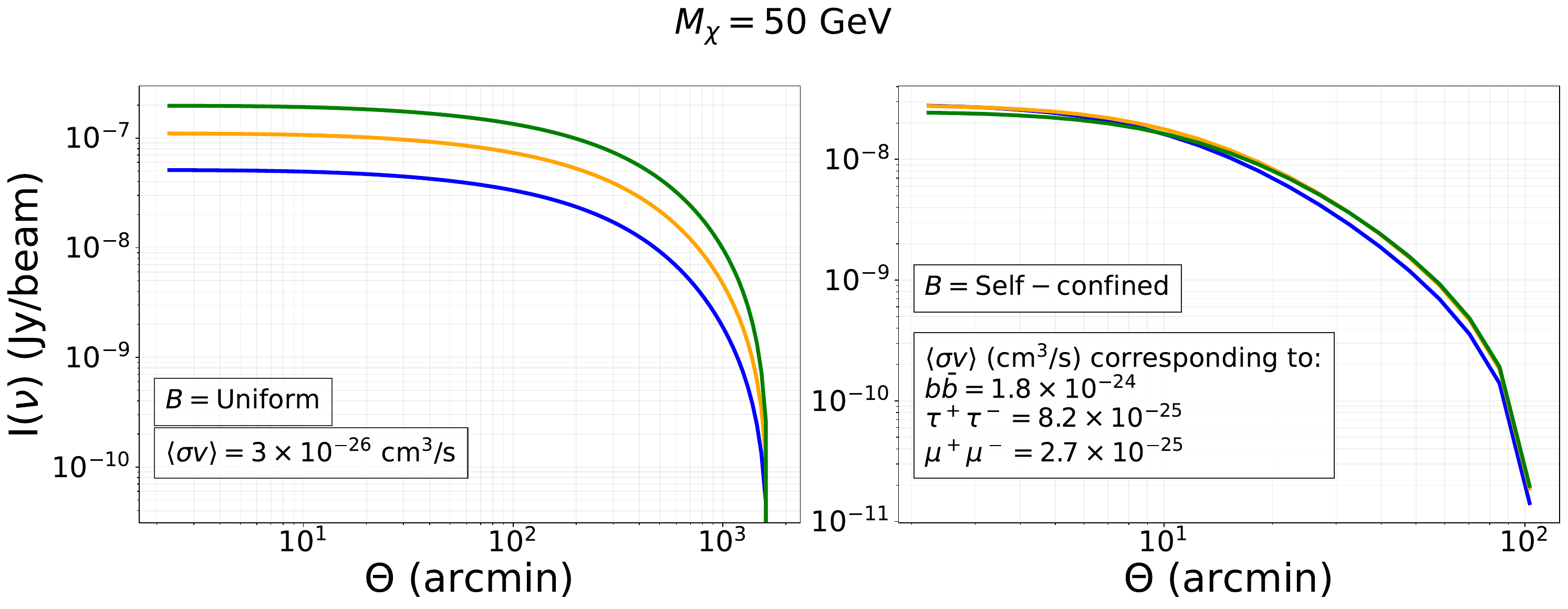}\\[0.4cm]
        \includegraphics[width=0.85\textwidth]{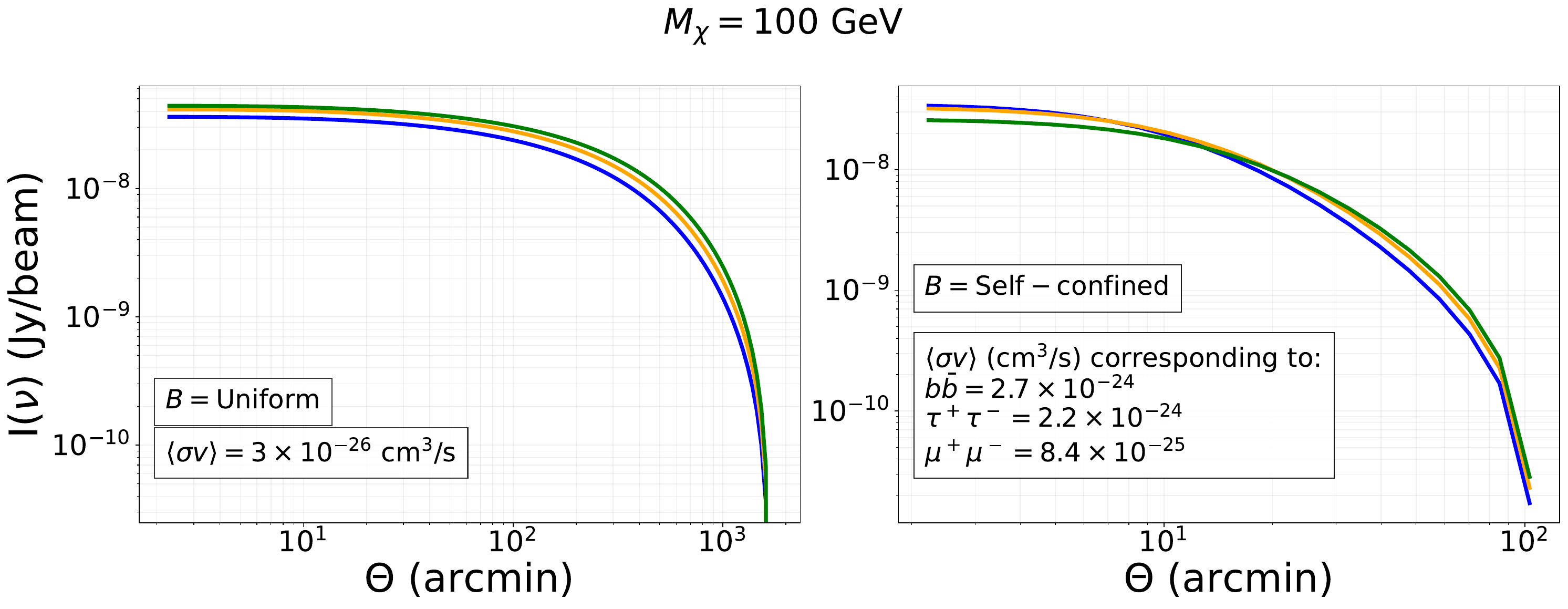}\\[0.4cm]
        \includegraphics[width=0.85\textwidth]{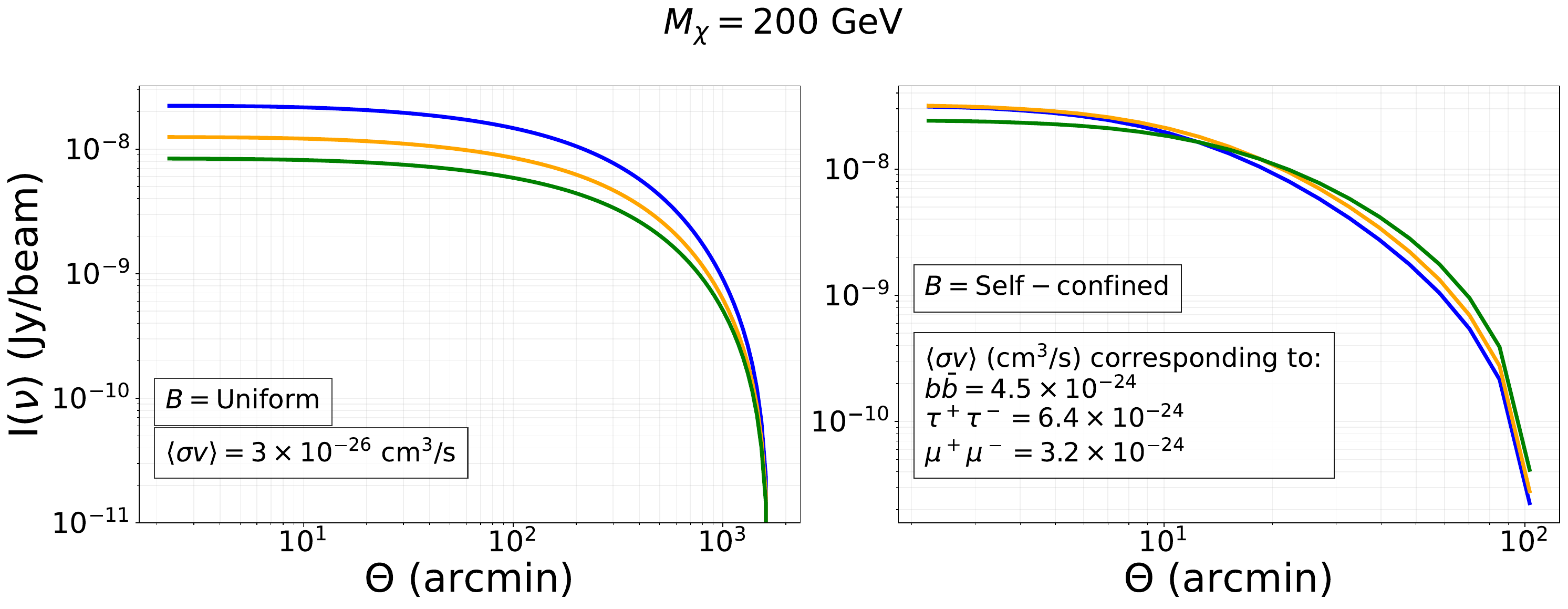}
    \end{minipage}
    \caption{Radial brightness profiles computed for a few WIMP masses annihilating into $b\bar{b}$, $\tau^+\tau^-$, and $\mu^+\mu^-$ assuming uniform (left), and self-confined (right) magnetic fields. The profiles for the self-confined magnetic field assumption were computed for cross-sections corresponding to the limit at that mass, as the cross-section determines the level of turbulence and thus the profile shape.}
    \label{fig:self-uniform}
\end{figure}

\begin{figure}[H]
        \centering
        \begin{minipage}{\textwidth}
        \centering
        \includegraphics[width=0.48\textwidth]{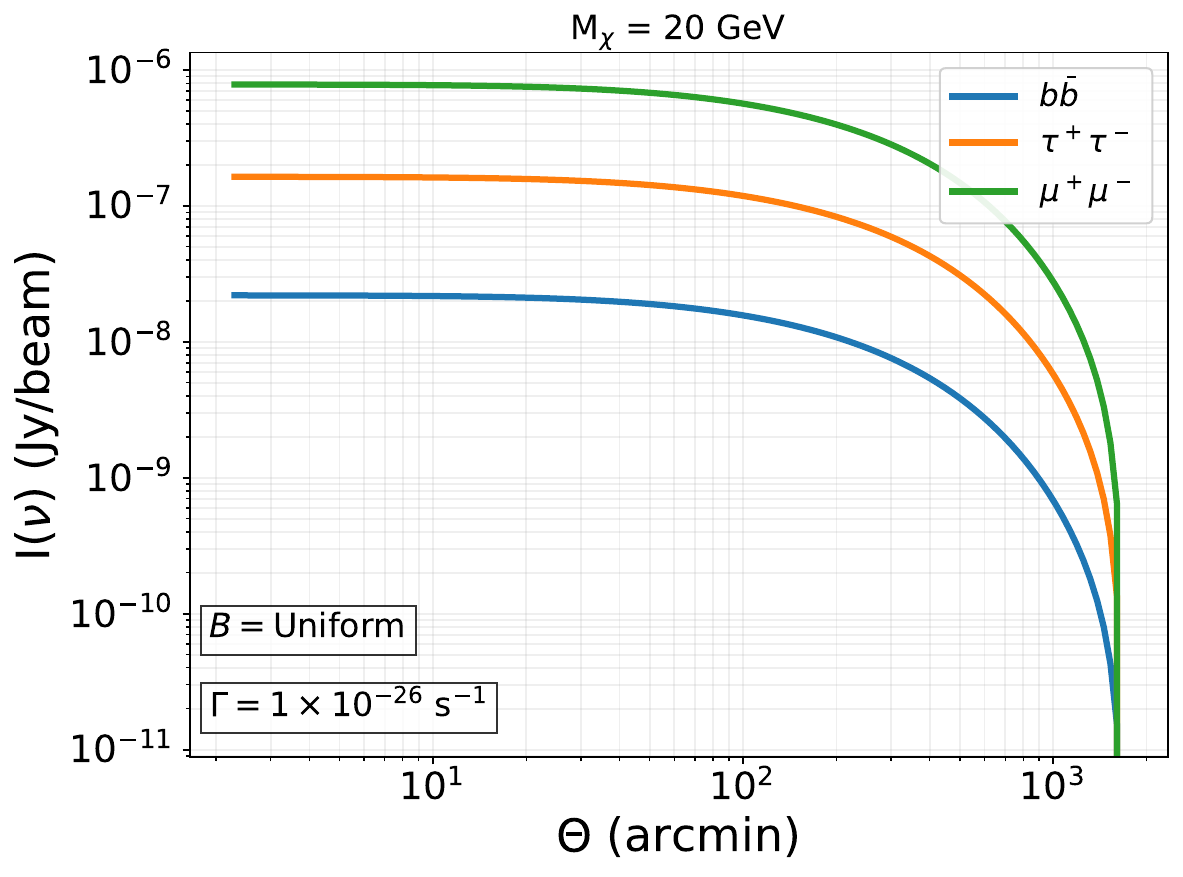}
        \includegraphics[width=0.48\textwidth]{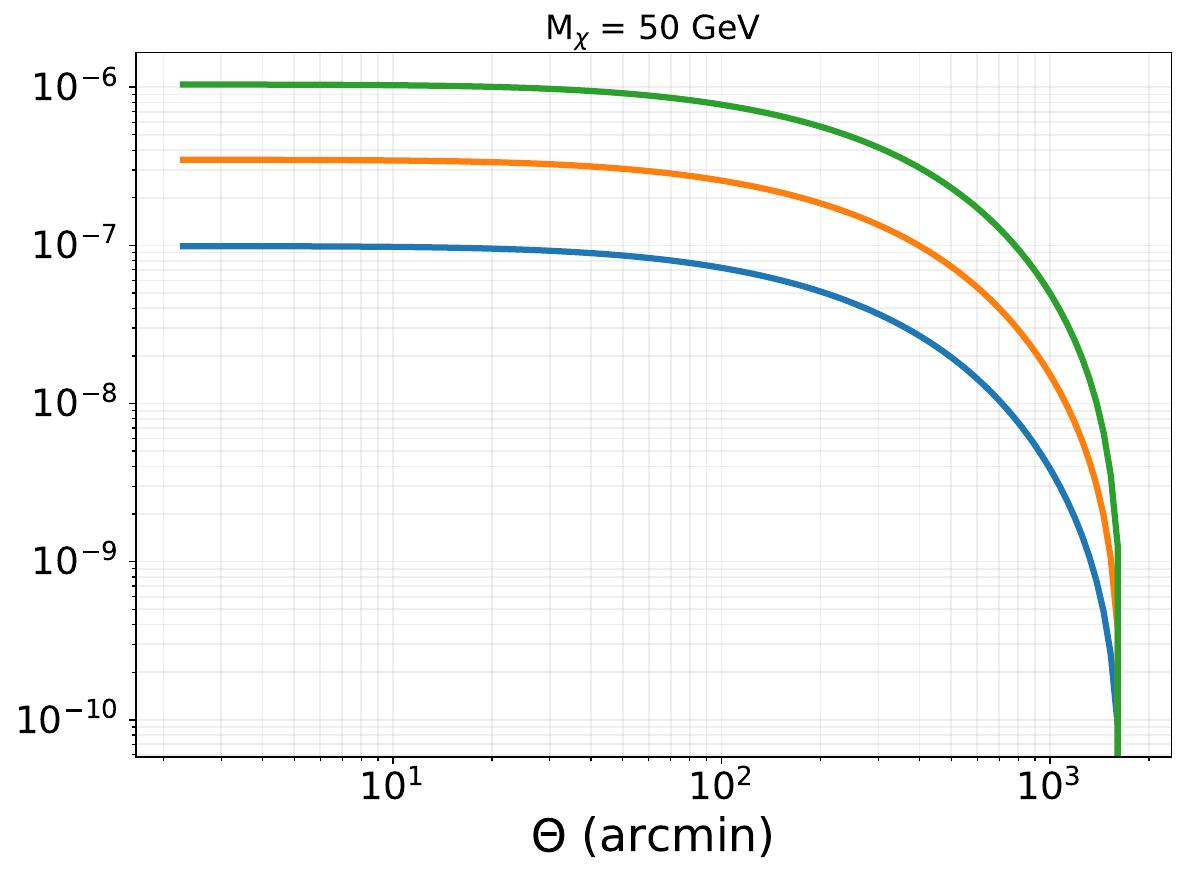}
        \includegraphics[width=0.48\textwidth]{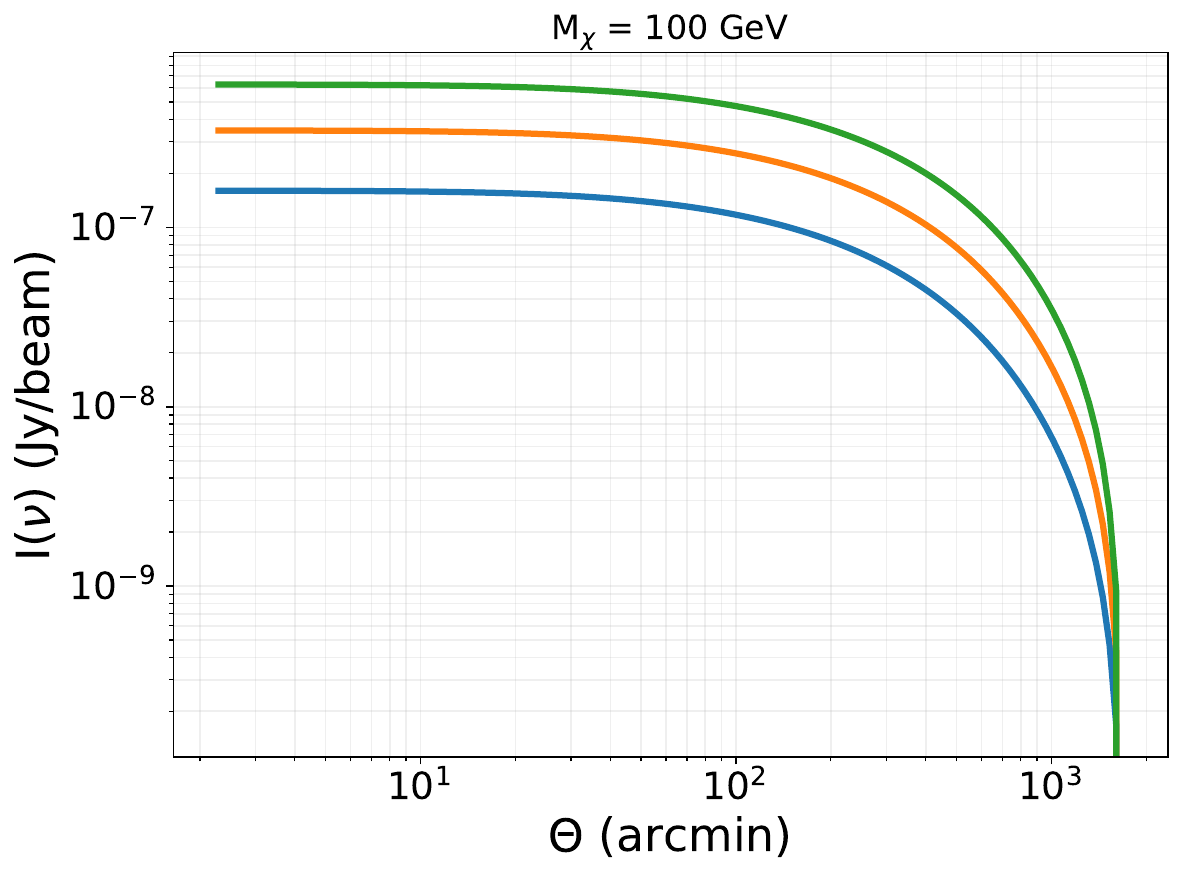}
        \includegraphics[width=0.48\textwidth]{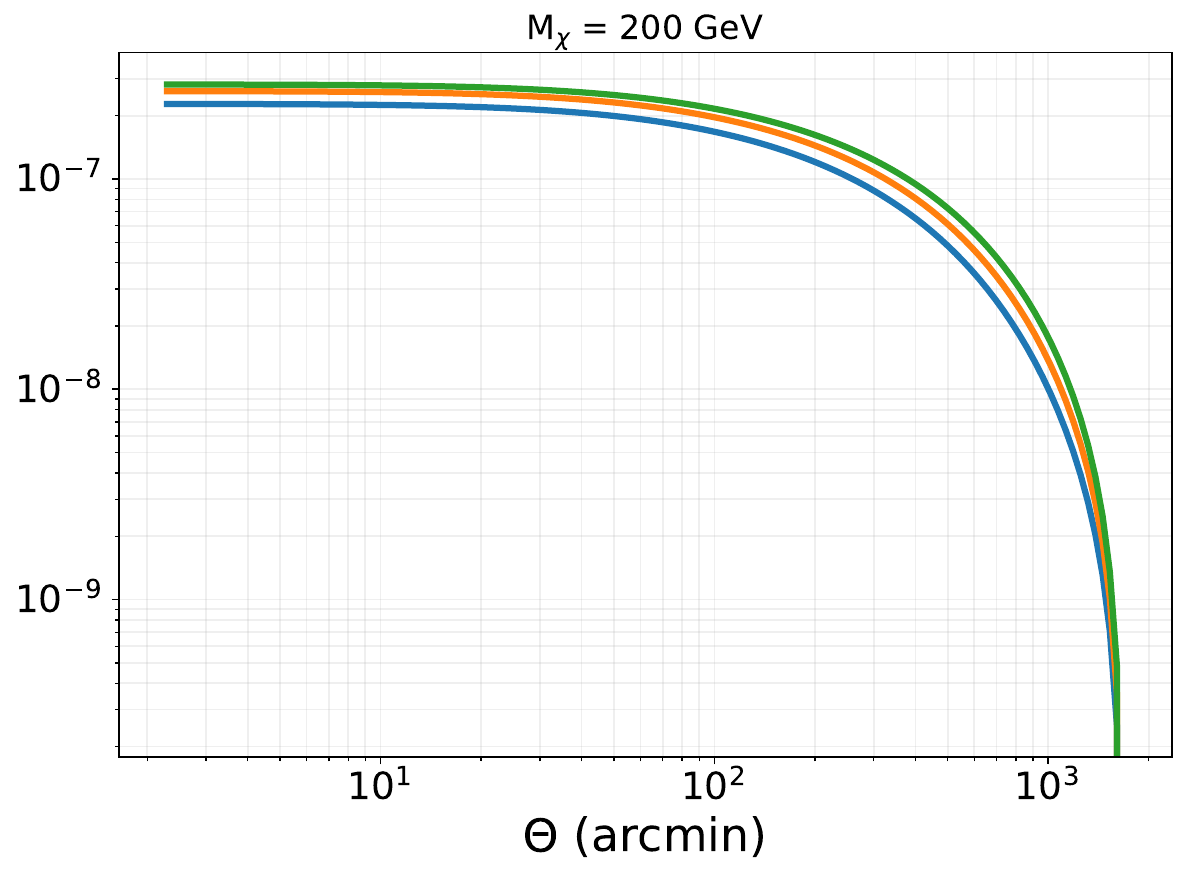}
    \end{minipage}
\caption{Radial brightness profiles computed for a few WIMP masses decaying into $b\bar{b}$, $\tau^+\tau^-$, and $\mu^+\mu^-$ assuming a uniform magnetic field.}
\label{fig:decayW}
\end{figure}

\section{MeerKAT Observations and Data Reduction}
\label{section:datardx}
\subsection{Observations of \textsc{Ret II}}

We conducted observations of \textsc{Ret II} using MeerKAT, a 64-dish array located in South Africa's Northern Cape, and a precursor to the SKA~\cite{jonas2016meerkat,meerkat_specs}, in the UHF band [544–1088 MHz].
\textsc{Ret II} was observed for 10 hours, corresponding to the allocated time for this target, alongside two calibrators: the primary calibrator \textsc{J0408-6545} used for flux density, delay, and bandpass calibration, and the secondary calibrator \textsc{J0303-6211}, for gain calibration. An integration time of 8 seconds was used for all observations, with the bandpass divided into 4096 channels.

\subsection{Data Reduction}

Data reduction is comprised of 3 main steps: flagging bad data, cross-calibration, and self-calibration. 
The observed visibility data contains corrupted measurements from the atmosphere, antenna systems, and receivers that must be corrected to accurately represent the sky.
To solve for these, calibration is performed using calibrator sources, which provide the correction factors needed to recover the true sky signal. 
We used the \texttt{CARACal} pipeline~\cite{jozsa2020caracal} for the flagging and cross-calibration.
Once we derived corrections from the calibrators, applied the corrections to our target, and flagged our target for known RFIs,
we apply the solutions to the target and proceed to self-calibration.

Self-calibration is where the target is used to iteratively improve the calibration solutions~\cite{pearson1984image}. This process adjusts the gains to minimize the difference between the observed visibilities and those predicted by the sky model. 
The self-calibration pipeline was orchestrated using
\texttt{Stimela}~\cite{makhathini2018advanced,smirnov2024africanus}, which allowed us to seamlessly integrate \texttt{WSClean}~\cite{wsclean} for imaging, \texttt{Quartical}~\cite{quartiCal} for calibration, and \texttt{Breizorro}~\cite{breizorro} for masking.
Self-calibration involved three rounds of calibration with multiple imaging steps. 
The pipeline begins with imaging the cross-calibrated data using \texttt{WSClean}. Standard imaging parameters were used with Briggs weighting (robustness 0.0)~\cite{briggs1995high,wsclean}. This initial image provides the starting sky model necessary for self-calibration. 
The first round of \texttt{QuartiCal} solves for direction-independent delay and offsets and phase gains.

After the first round of imaging the calibrated data, we iteratively imaged with an improved mask to find the optimum mask before moving on to the second round of calibration. 
To enable deep cleaning, the iterative masking and subtraction of compact bright sources down to the noise floor, we employed an iterative masking approach with \texttt{Breizorro}. With an initial high-threshold mask to only pick up the very bright sources. After the first clean, regions with artifacts were manually outlined using CARTA~\cite{carta} and given to~\texttt{Breizorro} as a regions file to exclude them.
After a few rounds, and once the image qualities had improved, the threshold was progressively lowered for subsequent masks, since the more the image improves, the less we are mistaking noise and artifacts for real sources. 
For the second round and any upcoming rounds, \texttt{QuartiCal} uses the improved sky model from the previous image. 

We inspected both the image and the residuals after each round of masking and imaging.
This iterative process continued until we were satisfied with the final mask and residuals. 
Once we image the corrected residuals, WSClean outputs both the residuals and the residuals of the residuals. We take the latter to \texttt{PyBDSF}~\cite{pybdsf} to remove persisting leftover sources.

\begin{figure}[H]
\centering
\begin{minipage}[t]{0.46\linewidth}
\centering
\includegraphics[width=\linewidth]{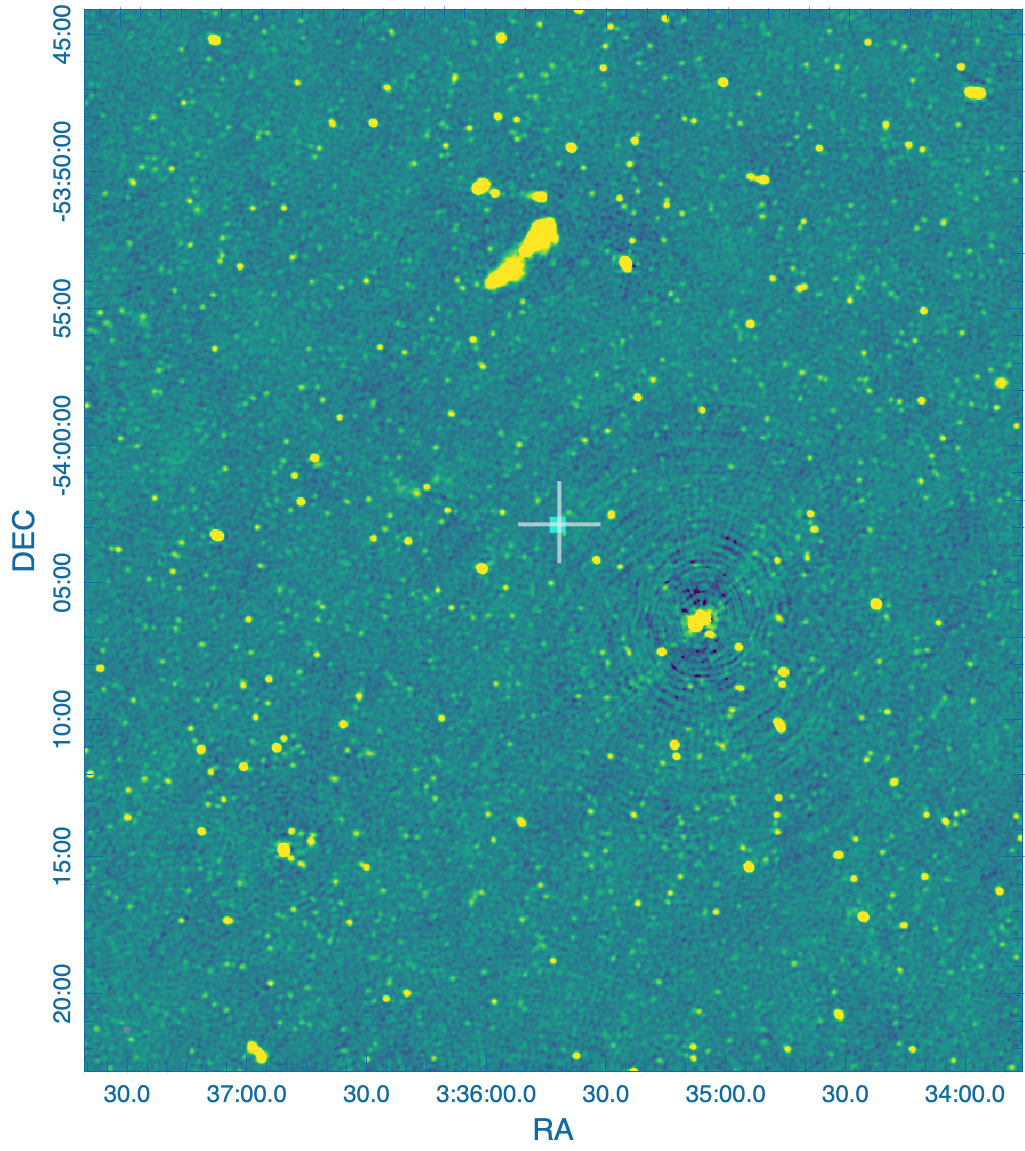}
\end{minipage}
\begin{minipage}[t]{0.485\linewidth}
\centering
\includegraphics[width=\linewidth]{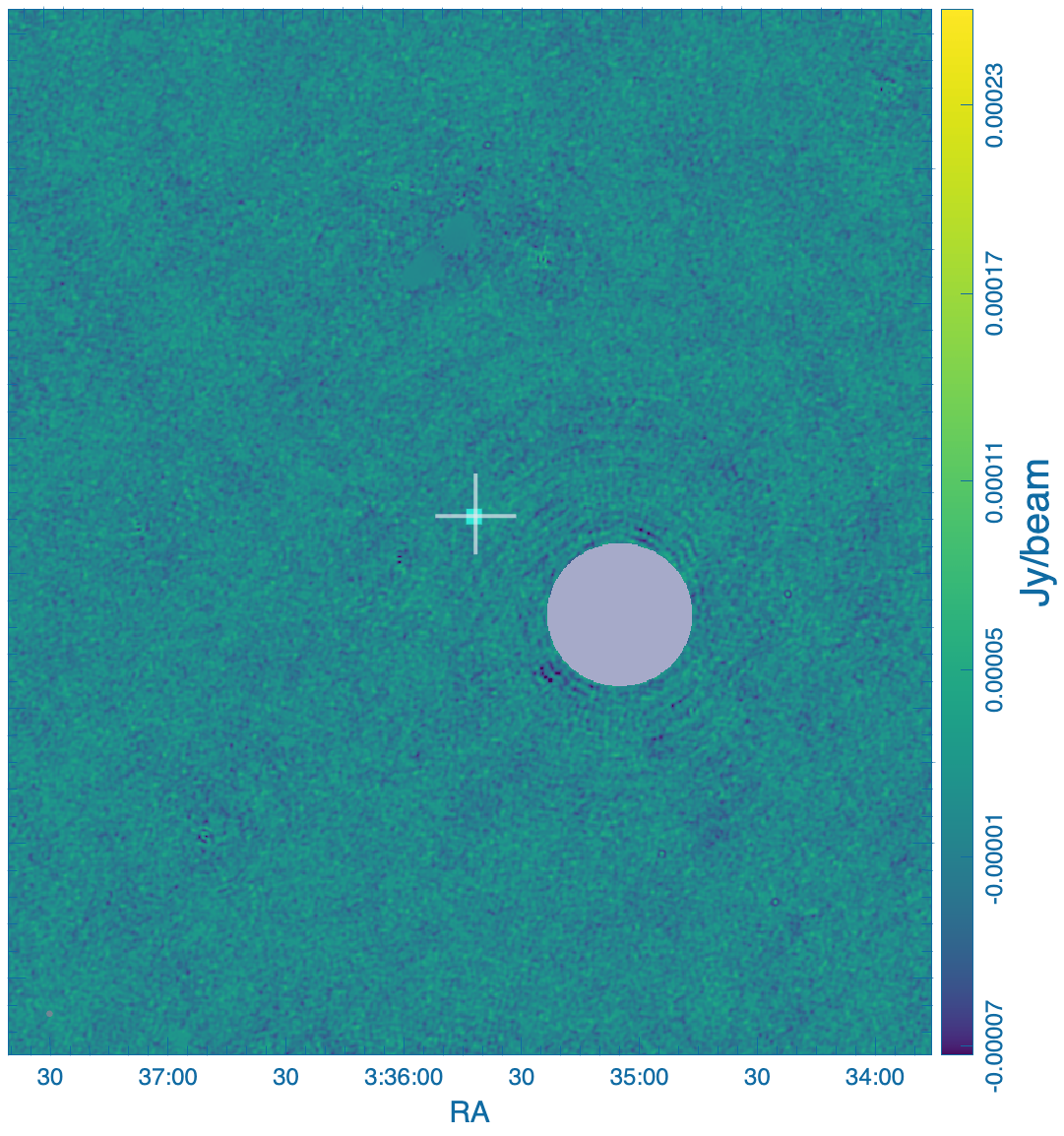}
\end{minipage}
    \caption{Final image of the \textsc{Ret II} field: Imaged using \texttt{WSClean} with Briggs (robust = 0) weighting (left) and \texttt{PyBDSF} residual map (final science image) (right). 
    The shaded, circled region is the excluded region due to poor calibration.}
    \label{fig:residual}
\end{figure}

\section{Discussion and Results}
\label{section:discres}
This section presents the upper bounds on DM annihilation and decay derived from our analysis of radio observations of Ret II. The discussion highlights the key factors influencing these constraints, beginning with the fundamental role of instrumental sensitivity and resolution. We then detail the statistical methodology used to establish the bounds and interpret their implications for DM particle properties.

\subsection{Sensitivity and Resolution}
The detectability of faint, diffuse radio emission from DM annihilation or decay is fundamentally limited by two instrumental properties: sensitivity and angular resolution. These parameters, set by the design and configuration of the interferometer, directly shape the robustness of any derived constraints.

Following the standard formalism of~\cite{thompson2017interferometry}, theoretical sensitivity $S_{\mathrm{rms}}$, for a Stokes I point source is given by the radiometer equation:
\begin{equation}
S_{\mathrm{rms}} = \frac{\text{SEFD}}{\sqrt{ 2\Delta \nu \,t_{\text{int}}\, N(N-1)}},
\label{eq:sensitivity}
\end{equation}
where, $N$ is the number of baselines, $\Delta \nu$ the observing bandwidth, $t_{\text{int}}$ the total integration time,
and the SEFD, the system equivalent flux density, which characterizes the combined sensitivity of the antenna and its receiving system~\cite{thompson2017interferometry} and is defined as 
$\text{SEFD} = \frac{2k T_{\text{sys}}}{A_{\text{eff}}}$
where $k$ is Boltzmann's constant, $T_{\mathrm{sys}}$ the system temperature, 
and $A_{\mathrm{eff}}$ the effective collecting area.

While this sensitivity equation is a simplified form from \cite{meerkat_specs} that assumes a natural weighting, it illustrates the fundamental scaling relationships for interferometric sensitivity.\\
The angular resolution is approximately
\begin{equation}
\theta \approx \frac{\lambda}{B_{\text{max}}},
\label{eq:resolution}
\end{equation}
where $\lambda$ is the observing wavelength and $B_{\text{max}}$ is the maximum baseline length.\\
High resolution is essential for separating diffuse emission from compact sources, ensuring that bright foreground objects can be identified and subtracted, leaving cleaner residual maps in which a potential DM signal can be identified and quantified.

In practice, sensitivity and resolution jointly define the quality of the visibility data. Short baselines enhance sensitivity to diffuse emission, while long baselines provide the resolution needed to isolate compact sources. The number of dishes, integration time, baseline distribution, and bandwidth establish the theoretical performance, but the outcome ultimately depends on calibration and imaging choices. 
Because the relative weighting of baselines determines the balance between sensitivity and resolution, finding an effective compromise is essential.\\
The trade-off between sensitivity and resolution in interferometric imaging arises from the different roles of baseline lengths: short baselines capture large-scale, diffuse emission, while long baselines provide high angular resolution. Imaging, therefore, requires weighting their contributions, with different schemes emphasizing one at the expense of the other. Common approaches include Briggs weighting \cite{briggs1995high} and tapering \cite{CASAdocs_Taper}.\\
In our analysis, a weighting with Briggs (robust = 0), was found to be optimal. Values towards -2 (more "uniform" weighting) suppress short baselines, favoring resolution at the cost of sensitivity to diffuse emission, while values towards +2 (more "natural" weighting) suppress the contributions from long baselines, enhancing sensitivity but reducing resolution. 
While Briggs weighting (+2) is more desirable in principle since it favors short baselines and maximizes sensitivity, it gives the worst resolution, making it difficult to do an accurate source subtraction.\\

Using the residual maps produced with the Briggs weighting (robust = 0), we compare the observed emission to the calculated dark-matter-induced synchrotron signals. These theoretical signals are computed for a range of WIMP masses, three annihilation/decay channels ($b\bar{b}$, $\tau^{+}\tau^{-}$, $\mu^{+}\mu^{-}$), and under the specific astrophysical assumptions outlined in Section~\ref{section:modeling}. The comparison is performed via a likelihood analysis (using a $\chi^{2}$ test statistic) to derive upper bounds on the annihilation cross-section $\langle \sigma v \rangle$ or decay rate ($\Gamma$) as a function of WIMP mass. 
We now detail the statistical methodology behind this approach.
\label{subsec:sen-res}
\subsection{Deriving Constraints via Likelihood Analysis}

\begin{figure}[H]
    \centering
    \begin{minipage}[t]{0.9\linewidth}
        \centering
    \includegraphics[width=\textwidth]{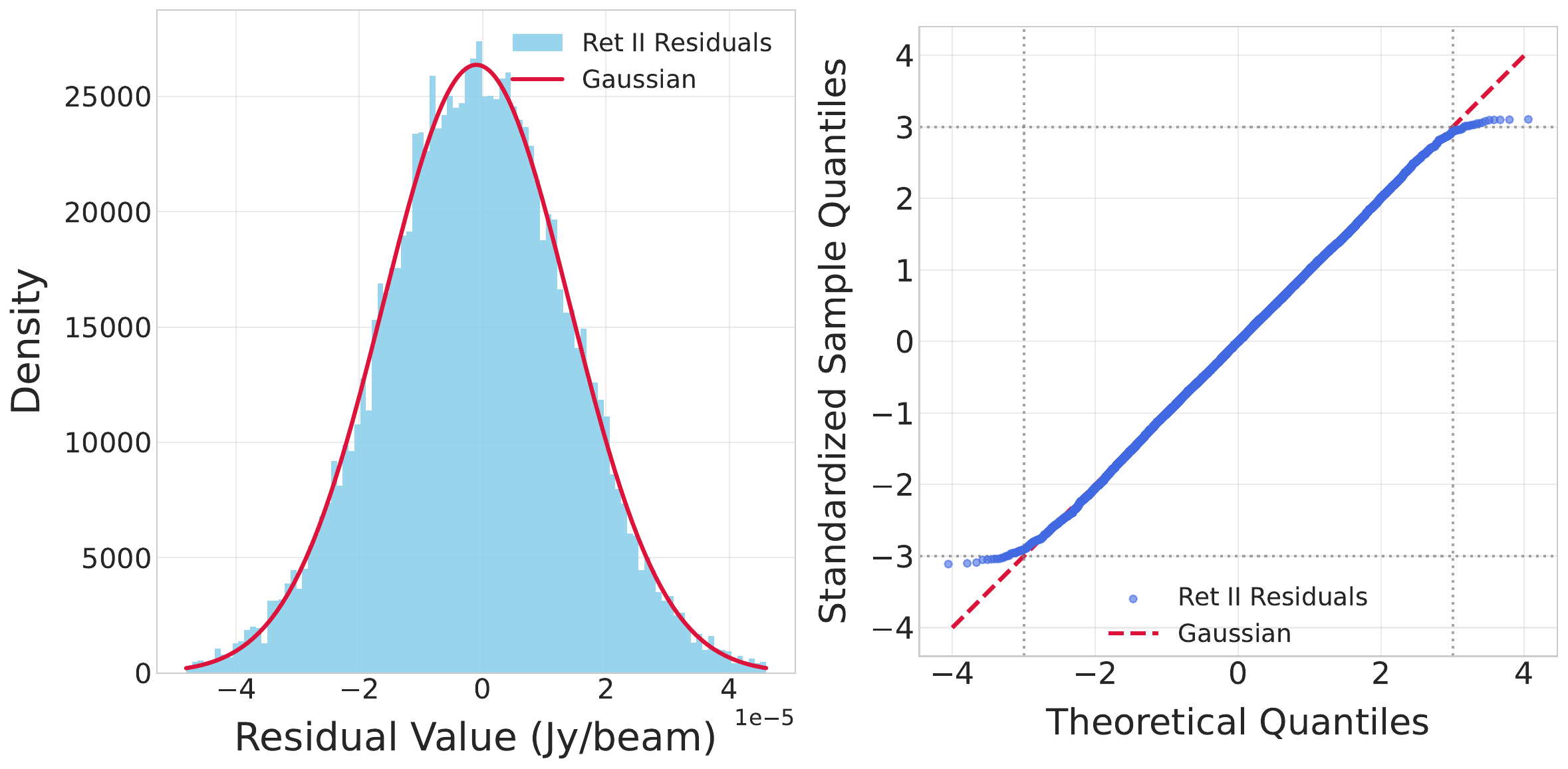}
        \par (Histogram - Q-Q plot)
    \end{minipage}
    \hfill
    \begin{minipage}[t]{0.9\linewidth}
        \centering
\includegraphics[width=\textwidth]{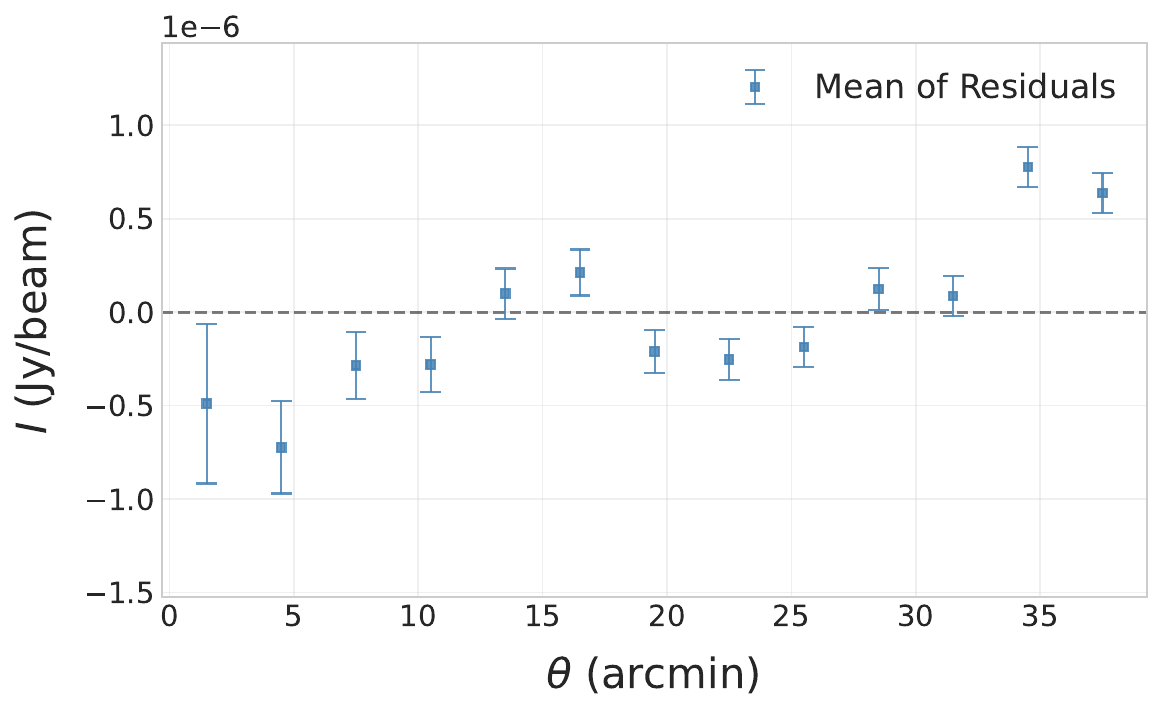}
    \par (Radial Brightness Profile)
\end{minipage}
\caption{Distribution of residual pixel values (top panel) and radial brightness profile (bottom panel) for \textsc{Ret II}. The plots illustrate the behavior of the residuals: the pixel values follow a Gaussian distribution centered on zero, and the radial profile is consistent with noise and showing no obvious extended emission attributable to DM.
Error bars represent the standard deviation of the mean 
($\sigma/\sqrt{N_{\rm beam}}$) where, $N_{\rm beam}$ is the number of independent beams in each annulus and $\sigma$ is the standard deviation in the annulus. }
\label{fig:gaussian_radial}
\end{figure}

To validate the appropriateness of our likelihood analysis, as a first diagnostic, we test whether the residual image is consistent with Gaussian noise. 
We use the graphical method to test Normality.
Fig.~\ref{fig:gaussian_radial} (top panel) shows the residual histogram with Gaussian fit and the Q-Q plot~\cite{andersen2018introduction}.
\textsc{Ret II} residuals closely follow a Gaussian distribution centered at zero. From the Q-Q plot we see that minor deviations occur beyond $\pm 3\sigma $, a regime that we exclude from our statistical analysis. 
To further check for the presence of any spatially extended emission potentially attributable to DM annihilation, we performed a radial brightness profile analysis.
We calculated the mean brightness of our region of interest from the centre outwards over concentric annuli, assuming spherical symmetry (following \cite{regis2017dark}).
We accounted for the spherical geometry of the sky by applying a declination correction in the RA direction: 
 $\Delta \text{RA}_{\rm true} = \Delta \text{RA} \cdot \sin\theta, \quad \theta = 90^\circ - \delta$,
which compensates for the convergence of RA lines at higher declinations.
This ensures that circular annuli in pixel coordinates correspond to true circles on the sky, preventing east-west squashing at higher declinations. The resulting flat brightness profile (Fig.~\ref{fig:gaussian_radial} bottom panel) shows no detectable excess emission that's attributable to an emission from DM.

We therefore derive upper bounds on the annihilation cross-section $\langle \sigma v \rangle$ and $\Gamma$ as a function of WIMP mass $M_\chi$.

To constrain $\langle \sigma v \rangle$ and $\Gamma$, we employ a likelihood ratio test.
Following the approaches in \cite{regis2017dark, sarkis2023radio}, we assume a Gaussian likelihood defined by:
\begin{equation}
    \mathcal{L} \propto e^{-\chi^2/2}, \quad 
    \chi^2 = \frac{1}{N_{\mathrm{bpix}}} \sum_{i=1}^{N_{\mathrm{pix}}} 
    \left( \frac{S_i^{\mathrm{th}} - S_i^{\mathrm{obs}}}{\sigma_i^{\mathrm{rms}}} \right)^2,
    \label{eq:chi2}
\end{equation}
where $N_{\mathrm{pix}}$ is the total number of pixels in the image, $N_{\mathrm{bpix}}$ is the number of pixels in one beam, $i$ denotes the pixel, $S_i^{\mathrm{th}}$ is the predicted synchrotron surface brightness from DM annihilations/decay convolved with the beam of our observed map, $S_i^{\mathrm{obs}}$ is the observed surface brightness in the residual map, and $\sigma_i^{\mathrm{rms}}$ is the local RMS noise. 
The sum in Eq.~\eqref{eq:chi2} is equivalent to:
\begin{equation}
 \sum_{j=1}^{N_{\mathrm{beam}}}\left(
    \frac{S_j^{\mathrm{th}} - S_j^{\mathrm{obs}}}
    {\sigma_j^{\mathrm{rms}}}
    \right)^2,
    \end{equation}
where $N_{\mathrm{beam}}$ is the number of independent beams.
%
Model images are generated using \texttt{DarkMatters}; once the surface brightness is computed, \texttt{DarkMatters} uses \texttt{astropy} to create a 2D pixel grid matching Ret II's coordinates. It then maps the surface brightness profile into the corresponding pixels to generate a 2D fits map. This is to enable a direct pixel-to-pixel comparison with the observed data. 
These model images (Jy/pixel initially) were reprojected to the pixel grid and coordinate system of the observed map to ensure consistent spatial sampling~\cite{reproject}. 
The reprojected models were then convolved with the synthesized beam of the observed map using \texttt{astropy} \texttt{Gassian2DKernel}~\cite{astropy2d} to convert the initially Jy/pixel units in the map to Jy/beam. \\
The sum runs over all $N_{\mathrm{pix}}$ pixels within the region of interest (ROI). The factor $N_{\mathrm{beam}}$ represents the number of independent beams within the ROI, calculated as the ratio of the ROI area to the effective area of the synthesized beam: 
\begin{equation}
    N_{\mathrm{beam}} = \frac{\Omega_\mathrm{ROI}}{\Omega_\mathrm{Beam}}.
\end{equation}
This weighting accounts for pixel correlation and ensures that we count independent beams. The effective area of the beam is given by:
\begin{equation}
    \Omega_\mathrm{Beam} = \frac{\pi}{4 \ln(2)} \cdot \mathrm{b_{maj}} \cdot \mathrm{b_{min}},
\end{equation}

where $\mathrm{b_{maj}}=10.6''$ and $\mathrm{b_{min}}=9.7''$ are the major and minor axes of the synthesized beam.
The $\chi^2$ statistic is computed iteratively for each annihilation channel and DM mass.
To mitigate the influence of outliers (leftovers from the source subtraction), we apply a mask that excludes pixels where the absolute value of the surface brightness exceeds $3\sigma_{\mathrm{RMS}}$ the RMS noise (i.e., excluding both very positive and very negative fluctuations). \\
The 95\% confidence level (CL) upper limit on $\langle \sigma v \rangle$ for each mass and annihilation channel is determined by increasing $\langle \sigma v \rangle$ from its best-fit value (that we found consistent with no signal) until the fit becomes significantly worse. More precisely, we exclude cross-section values where,
\begin{equation}
    \Delta \chi^2 = \chi^2(\langle \sigma v \rangle) - \chi^2_{\mathrm{min}} > 2.71
\end{equation}
for a one-sided limit \cite{sarkis2023radio}, where $\chi^2_{\mathrm{min}}=\chi^2(\langle \sigma v \rangle_{bf})$ with $\langle \sigma v \rangle_{bf}$ being the best-fit cross-section (in our case it is essentially zero since there is no evidence for a signal).  
We use the same statistical procedure to derive bounds for decaying DM, by replacing $\langle \sigma v \rangle$ with $\Gamma$ in above considerations.
\subsection{Constraints on WIMP Properties}
\subsubsection{Annihilating WIMPs}
\begin{figure}[H]
    \centering
    \begin{minipage}[t]{0.49\linewidth}
        \centering
\includegraphics[width=\linewidth]{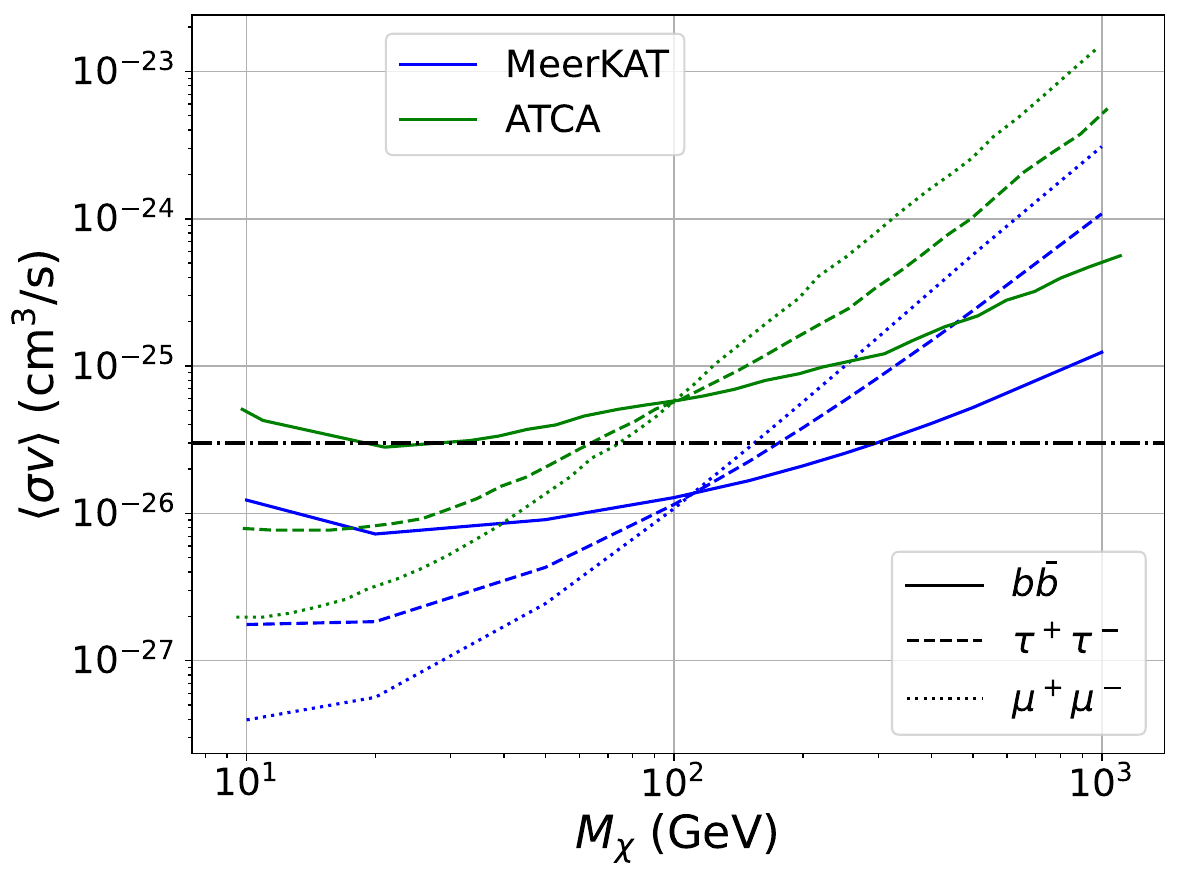}
    \end{minipage}
   \begin{minipage}[t]{0.49\linewidth}
        \centering
\includegraphics[width=\linewidth]{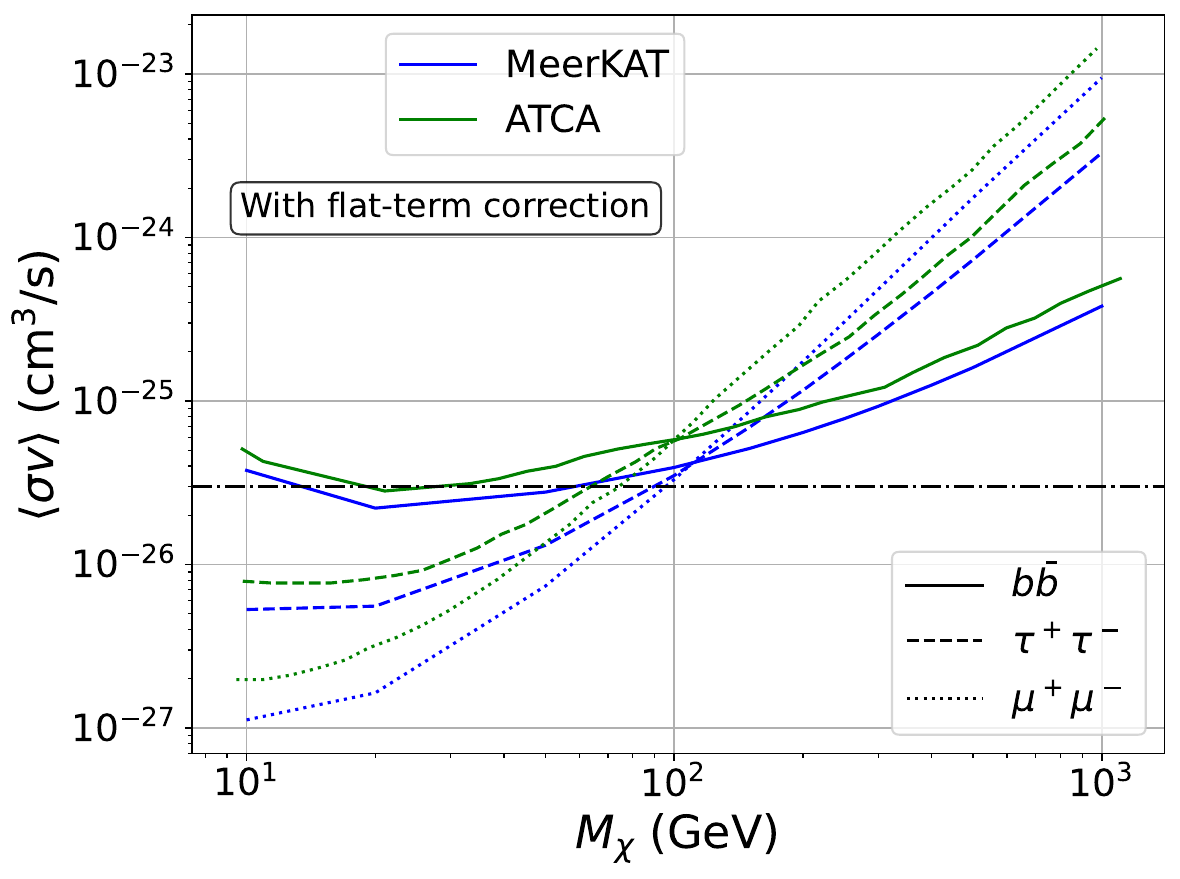}
\end{minipage}
\begin{minipage}[t]{0.49\linewidth}
    \centering
    \includegraphics[width=\linewidth]{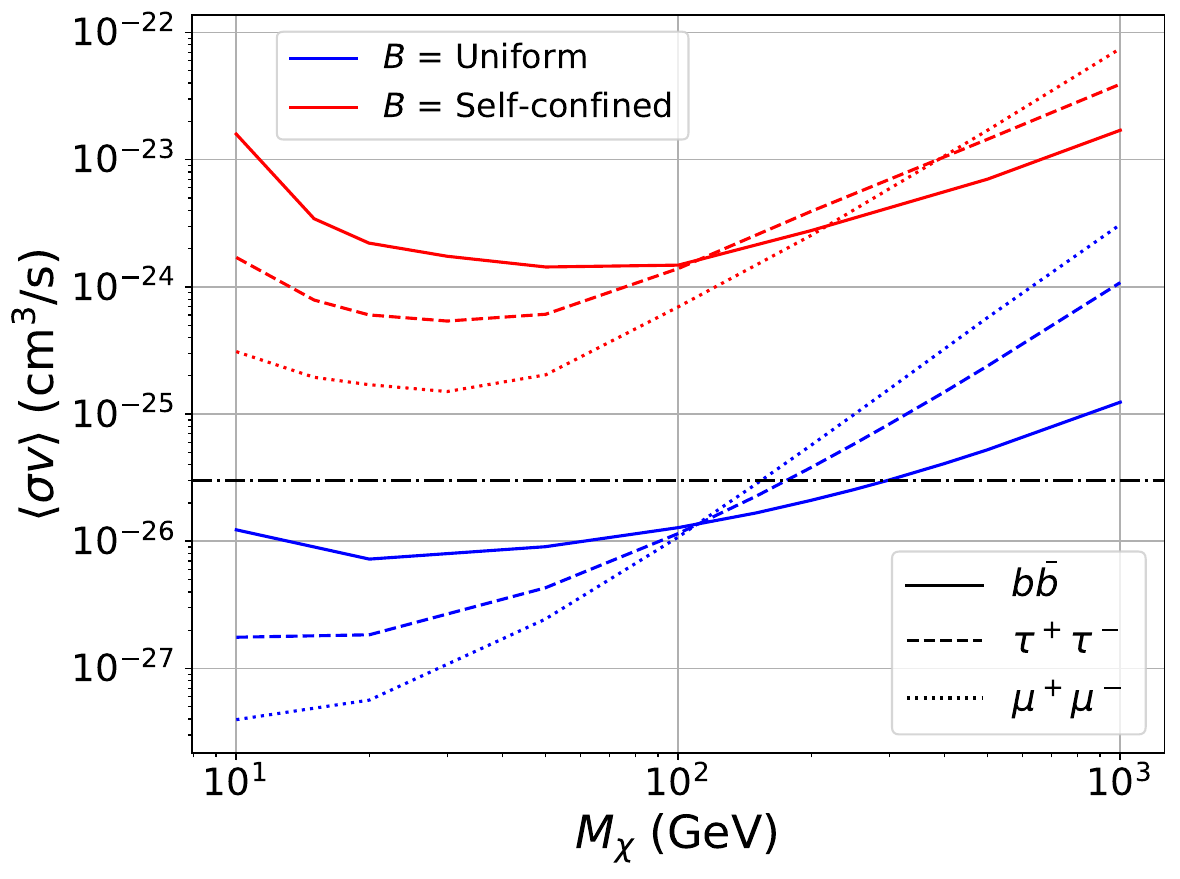}
\end{minipage}
    \begin{minipage}[t]{0.49\linewidth}
    \centering
    \includegraphics[width=\linewidth]{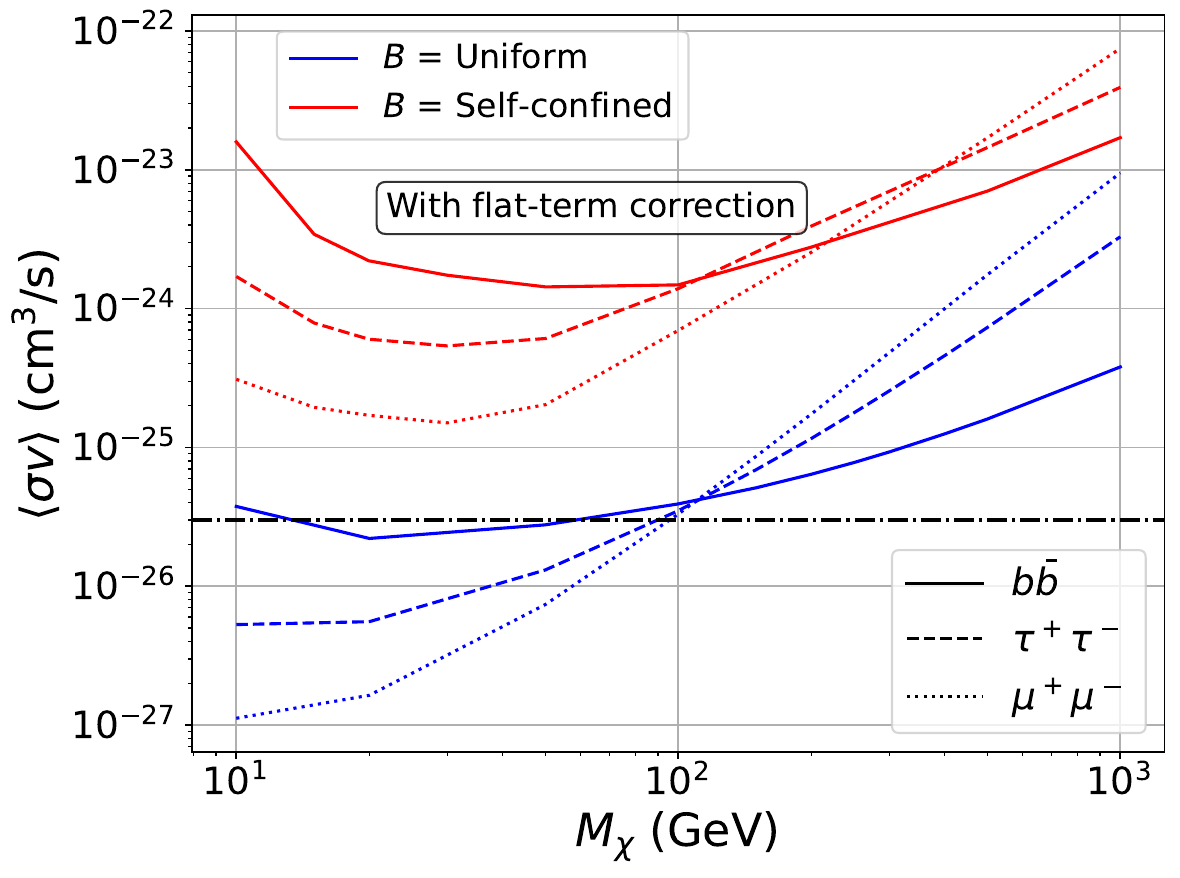}
\end{minipage}
\caption{Upper bounds (95\% CL) on the velocity-averaged annihilation cross-section 
    $\langle \sigma v \rangle$ for the $b\bar{b}$, $\tau^+\tau^-$, and $\mu^+\mu^-$ channels. 
     Top panels show upper bounds under the uniform magnetic field assumption compared with previous bounds from ATCA observations of \textsc{Ret II}~\cite{regis2017dark}. 
    Bottom panels show the comparison of the upper bounds computed with
    a spatially uniform magnetic field and diffusion (cf. Fig.~\ref{fig:self-uniform} left panel), with the bounds computed in the self-confinement scenario (cf. Fig.~\ref{fig:self-uniform} right panel).
    Left (right) panels show the bounds without (with) the inclusion of a spatially flat-term correction discussed in the text.
    The horizontal dashed line marks the benchmark annihilation cross section ($3 \times 10^{-26}\ \text{cm}^3/\text{s}$).}
    \label{fig:ULs_comp} 
\end{figure}

\begin{figure}[H]
    \centering
    \begin{minipage}[t]{0.325\textwidth}
        \centering
        \includegraphics[width=\textwidth]{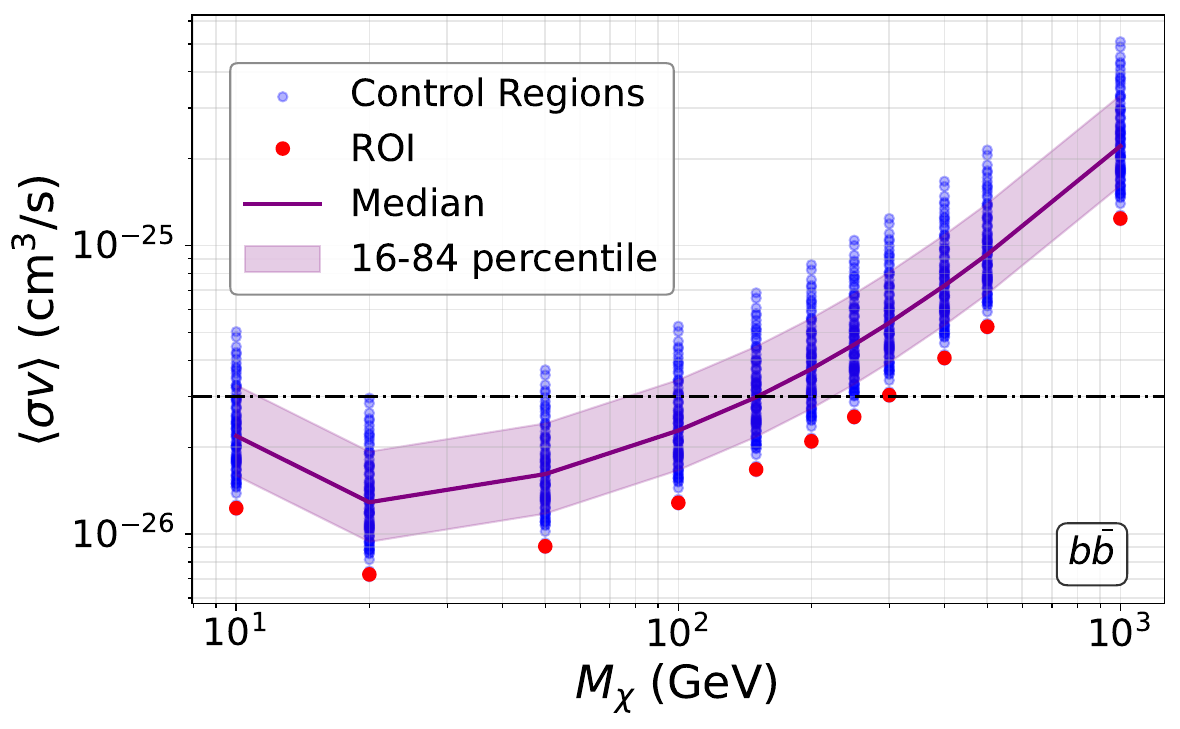}     
    \end{minipage}
    \hfill
    \begin{minipage}[t]{0.325\textwidth}
        \centering
        \includegraphics[width=\textwidth]{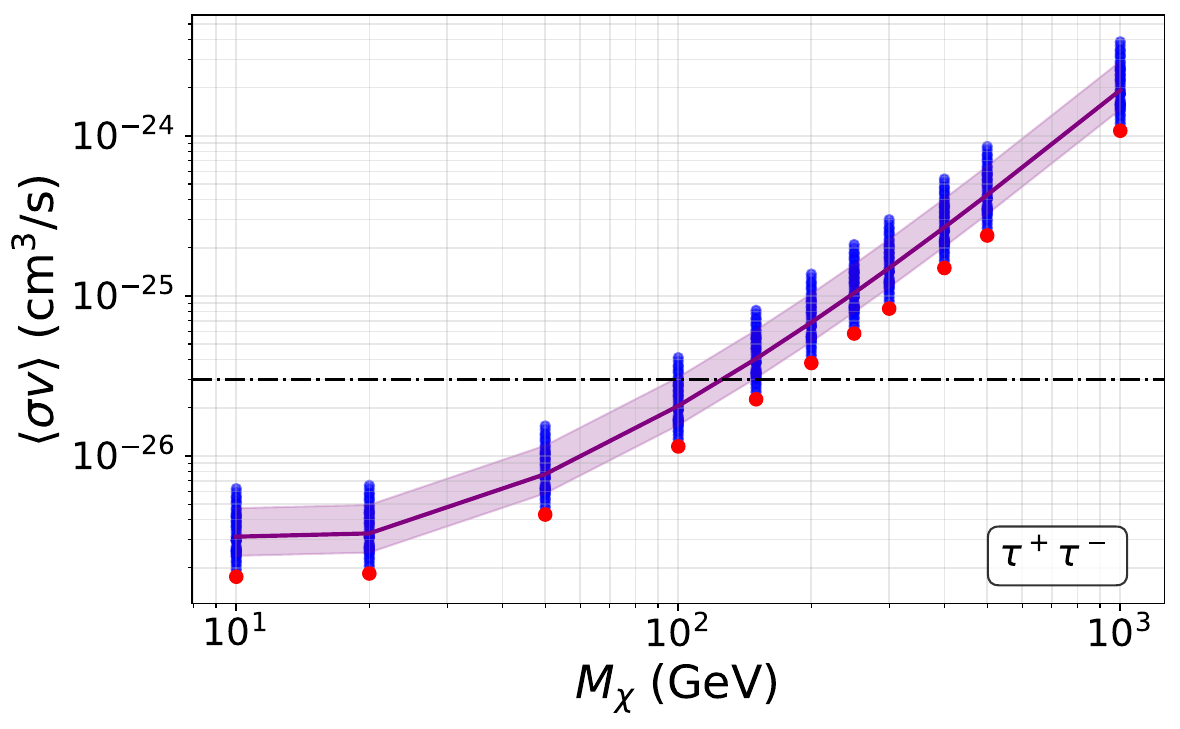}
    \end{minipage}
    \hfill
    \begin{minipage}[t]{0.325\textwidth}
        \centering
        \includegraphics[width=\textwidth]{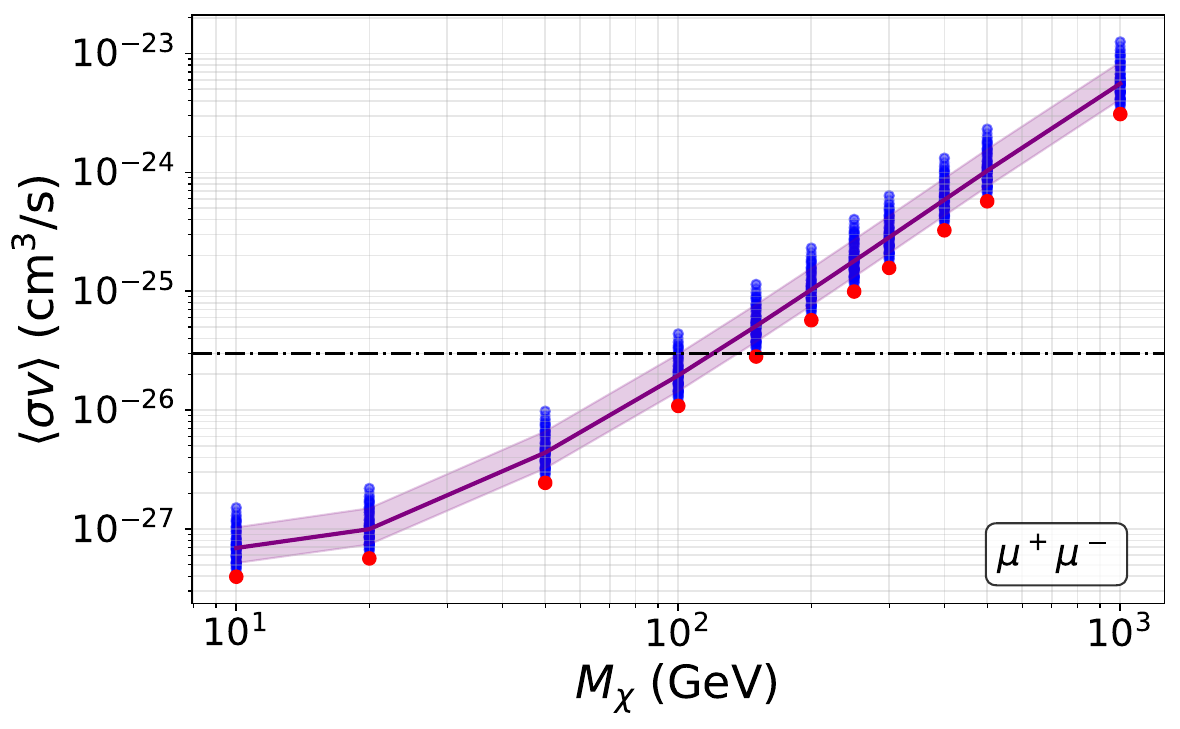}
    \end{minipage}
    \caption{Upper bounds for the $b\bar{b}$, $t\bar{t}$ and $\mu^+\mu^-$ channels for a few randomly produced control regions in the \textsc{Ret II} image field, including the central ROI region.}
    \label{fig:control-ULs}
\end{figure}

For WIMP DM, the typical target of indirect searches is the so-called thermal relic cross-section, corresponding to the value $\langle \sigma v \rangle$ expected if WIMPs froze out from thermal equilibrium in the early universe to yield the observed DM density today.
The surface brightness profiles in Fig.~\ref{fig:self-uniform} show that the expected WIMP emission reaches a level exceeding the noise of our observations for low mass WIMPs and thermal cross section, in the optimistic scenario of uniform magnetic field.
In Fig.~\ref{fig:ULs_comp}, we show that, in such a scenario, our analysis excludes annihilation cross-sections above the thermal value for masses 10–100 GeV, and regardless of the channel of annihilation.
Note that bounds for the $b\bar{b}$, $\tau^+\tau^-$, and $\mu^+\mu^-$ channels have different scalings. To understand this, we need to remember that for the radio frequency and magnetic field discussed here, the energy of the emitting $e^\pm$ is around 10 GeV. For low DM masses, those $e^\pm$ are more efficiently produced if the spectrum is harder (as in the case of $\tau^+\tau^-$, and more of $\mu^+\mu^-$), whilst if the DM mass is large, they are more copiously produced by a softer spectrum (as for $b\bar{b}$).
In Fig.~\ref{fig:ULs_comp} (top), we show that our constraints surpass previous ATCA's results for \textsc{Ret II}~\cite{regis2017dark}.
This is due to the improvement in the RMS sensitivity  
and, for small DM masses, to the lower frequency of MeerKAT observations (0.54-1.1 GHz versus 1.1-3.1 GHz). 
In the bottom panels of Fig.~\ref{fig:ULs_comp}, the bounds obtained in the optimistic scenario of a uniform magnetic field (blue lines) are compared to the more conservative case of self-generated turbulence (red lines).
The latter scenario leads to DM bounds that are around two orders of magnitude weaker than in the optimistic case for high DM masses, while the ratio increases at low masses up to about three orders of magnitude for $M_\chi\simeq10$ GeV. This increase is due to the low strength chosen for the coherent magnetic field $B_0=0.1\,\mu$G. Indeed, while at high DM masses the signal is nearly independent from $B_0$, since its effects in the synchrotron power and in the diffusion coefficient approximately cancel~\cite{regis2023self}, this is not the case for low masses and low $B_0$. This is because a low $B_0$ increases the energy of $e^\pm$ required to efficiently produce synchrotron emission at a given frequency, something which becomes harder and harder to achieve as the WIMP mass decreases.\\
In addition to our ROI, which is centered at the dynamical center of \textsc{Ret II}, we perform a likelihood analysis on multiple control regions centered at different locations to verify the uniformity of our residual map. 
We use the same template of the DM model for all the control regions, since the DM signal would decrease radially from the center, whilst we aim at testing the sensitivity uniformity of our observations.
The goal is to test whether different regions of the field exhibit similar characteristics against an identical model signal template. If residuals exhibit uniform behavior across different regions in the field, then we expect a similar chi-square distribution. 
Control regions were randomly located and have the same size as the ROI. 
The field size used is $\approx 84\times84$ arcminutes ($2048 \times 2048 $ pixels, with an edge buffer of 124 pixels excluded from the analysis).
The central region provides better DM bounds than the other regions, see Fig.~\ref{fig:control-ULs}. Our central region is slightly more sensitive than the rest because it has an optimal primary beam response and minimal calibration errors and, in turn, represents our best possible ROI; however, the overall RMS and signal-to-noise are not significantly better than the rest of the field. The improvement in the bounds can be actually explained by the slightly negative mean in the central region, which is very minor (at a level smaller than a few percent of the RMS noise, see distribution in Fig.~\ref{fig:gaussian_radial}), but constrains positive contribution, given the large number of independent beams in the statistical analysis.
To overcome this issue, we add a spatially flat term to the theoretical model, in order to account for possible systematic offset in the map.
We found that such term is always preferred over the DM contribution in the likelihood analysis, essentially because of the negative sign of the mean, with no indication of a non-zero DM contribution. 
Therefore the flat term can be simply computed as:
$\sum_i (S_i / \sigma_i^2)/\sum_i (1 / \sigma_i^2)$, 
where $S_i$ is the observed surface brightness in pixels $i$ and $\sigma_i$ is the RMS noise.
In Fig.~\ref{fig:ULs_comp}, we report bounds from the central ROI both without (left panels) and with (right panels) flat-term correction applied.
To test the expected statistical behavior under noise-dominated conditions, we perform a likelihood analysis on simulated residuals.
The simulated residuals represent an idealized scenario where systematics and remnants from sources are absent.
The goal is to establish a baseline for upper bounds under ideal conditions by creating simulated residuals from pure Gaussian noise convolved with the PSF of the real residual.
The simulated residual map is normalized to the minimum RMS of the residual map, which is 7.86~$\mu$Jy~beam$^{-1}$.
We then produce upper bounds from this map and compare it with our real data bounds. 
We can notice that in the case without the addition of the flat term, our bounds are significantly more optimistic than in the simulation. On the other hand, after the addition of the flat term, the agreement is remarkable, with the actual bounds being only slightly more conservative than in the idealized case of the simulation.
\begin{figure}[H]
    \centering
    \begin{minipage}[t]{0.48\textwidth}
        \centering
        \includegraphics[width=\textwidth]{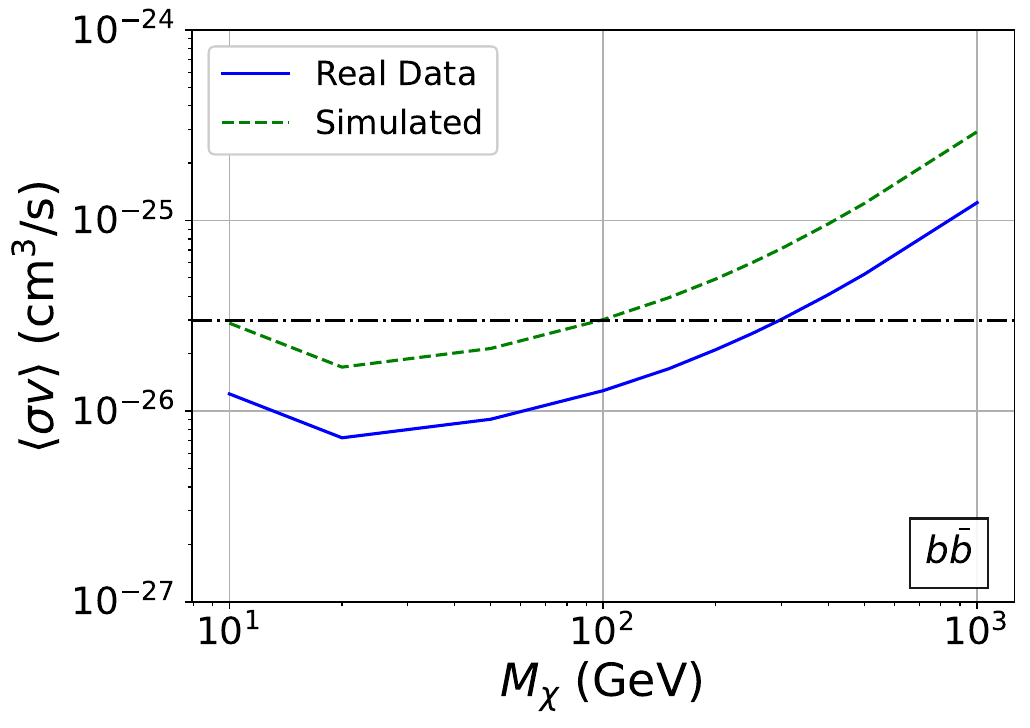}     
    \end{minipage}
    \hfill
    \begin{minipage}[t]{0.48\textwidth}
        \centering
        \includegraphics[width=\textwidth]{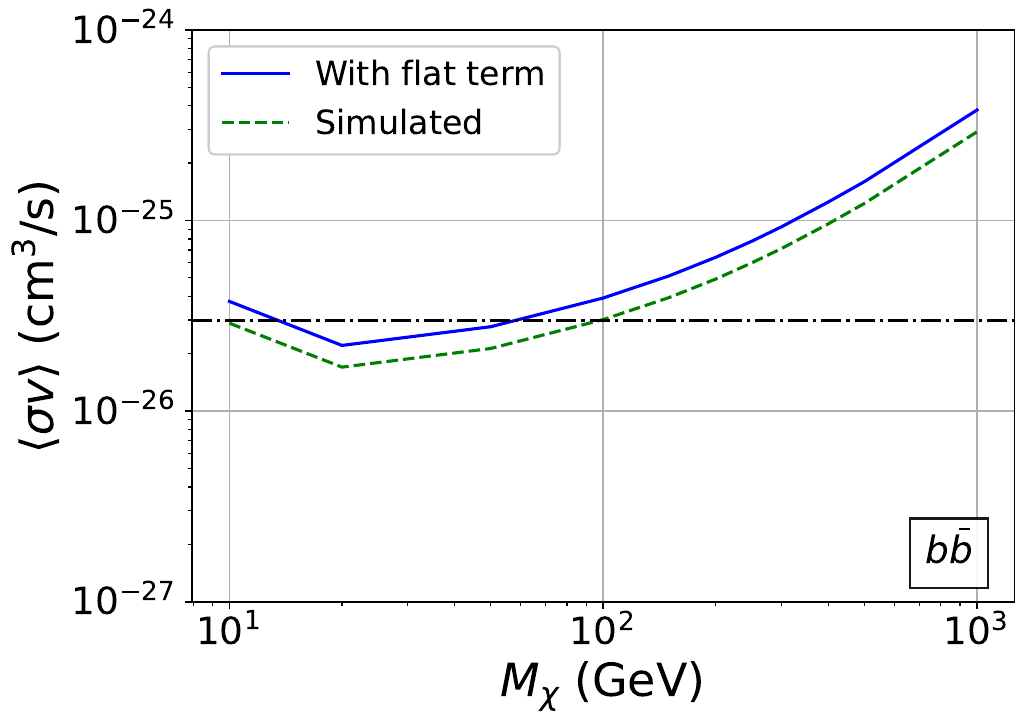}
    \end{minipage}
    \caption{Upper bounds from the simulated residual map with RMS values of 7.86~$\mu$Jy~beam$^{-1}$, compared to the upper bounds derived from the real data (with and without the flat term), for a WIMP annihilating into the $b\bar{b}$ channel.}
    \label{fig:sim-bounds}
\end{figure}
\subsubsection{Decaying WIMPs}

\begin{figure}[H]
    \centering
    \includegraphics[width=0.8\textwidth]{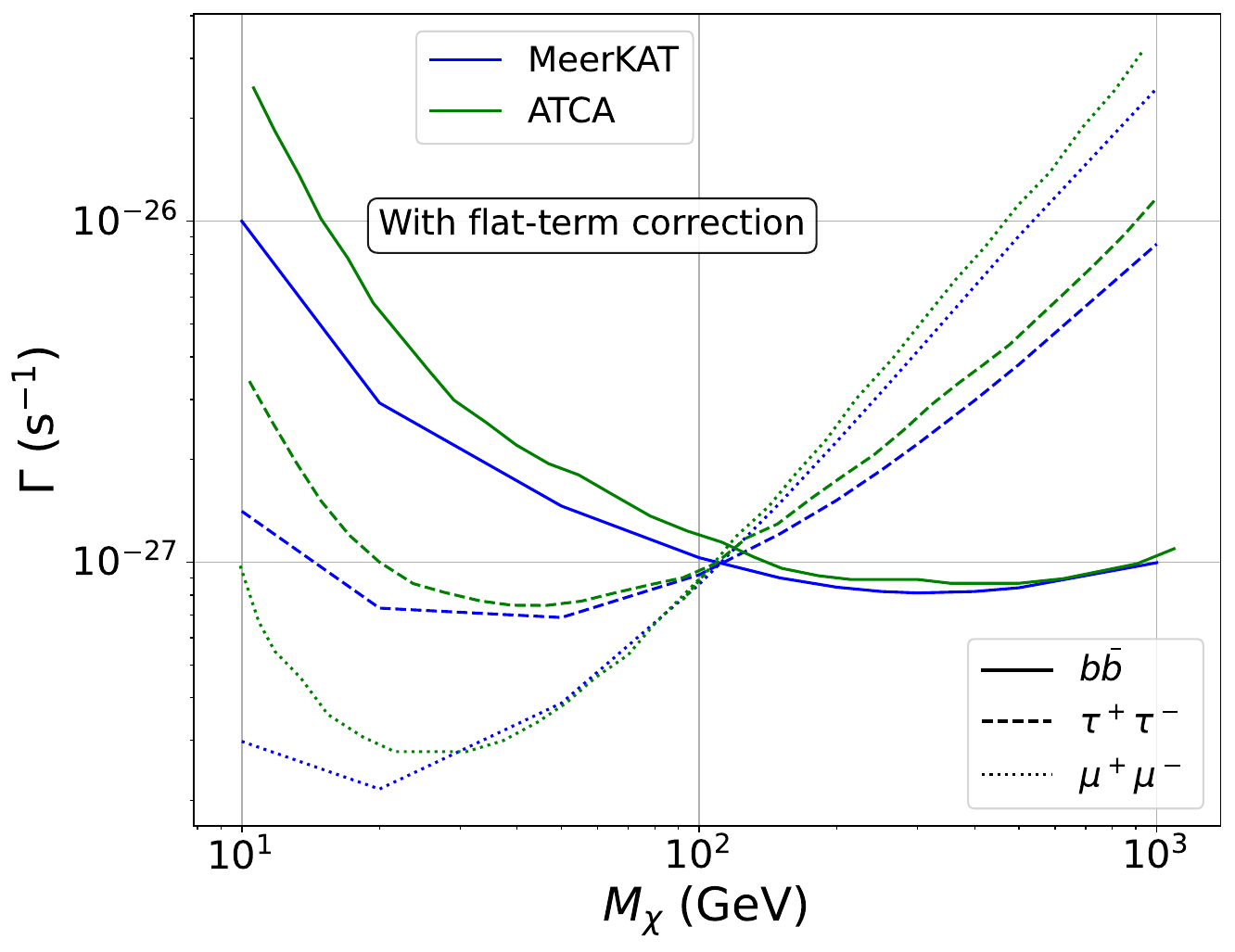}
    \caption{Upper bounds (95\% CL) on the decay rate  $\Gamma = \tau^{-1}$
    as a function of WIMP mass for the $b\bar{b}$, $\tau^+\tau^-$, and $\mu^+\mu^-$ 
    channels (with flat term corrections applied), under the uniform magnetic field assumption (cf. Fig.~\ref{fig:self-uniform} left panel), compared with \textsc{Ret II} ATCA's upper bounds from~\cite{regis2017dark}.}
    \label{fig:ULs_decay}
\end{figure}

For decaying DM, there is no analogous reference value for the decay rate ($\Gamma$) as the thermal cross-section for the annihilating scenario.
Clearly the DM must be significantly longer-lived than the age of the universe; otherwise, its abundance would have already decreased to levels inconsistent with cosmological observations. Our bounds on the lifetime are orders of magnitude longer than the age of the universe.
Surface brightness profiles for the decaying WIMP are shown in Fig~\ref{fig:decayW}.
Following the same statistical approach as for the annihilation analysis, we derive upper bounds on the decay rate ($\Gamma$) for the $b\bar{b}$, $\tau^+\tau^-$, and $\mu^+\mu^-$ channels, in the optimistic uniform scenario for the magnetic field. 
Results are shown in Fig.~\ref{fig:ULs_decay}, together with a comparison with previous results obtained with ATCA observations of \textsc{Ret II}~\cite{regis2017dark}.

\section{Conclusions}
\label{section:conc}
In this work, we presented results from observations of \textsc{Ret II}, a nearby, DM-rich dSph, using the MeerKAT radio telescope. The target was observed for 10 hours in the UHF band [544–1088 MHz]. After a rigorous data reduction and source subtraction, we obtained a residual map with an RMS sensitivity reaching 7.86 $\mu$Jy/beam, suitable for DM analysis and, in general, to search for diffuse emission from \textsc{Ret II}.  

No significant excess emission above the RMS noise level was detected in the residual map, and therefore possible DM-induced signals are fainter or comparable to the RMS of our observed data. 
This null result is consistent with previous gamma-ray and radio searches \cite{zhao2018searching, hoof2020global, regis2017dark}, including the disputed claim in \cite{geringer2015indication}. 
The DM emission we investigate is associated with synchrotron radiation from $e^\pm$ injected by DM annihilation or decay.
The main theoretical systematic uncertainties are related to the magnetic field assumptions, which play a crucial role in predicting the residence time of $e^\pm$ and the synchrotron emission. 

We considered three magnetic field scenarios: spatially uniform Milky-Way-like diffusion and magnetic field strength of $1\,\mu$G (optimistic), a $0.1\,\mu$G coherent magnetic field with turbulence self-generated by DM~\cite{regis2023self}, and a magnetic field confined within the half-light radius of the stellar region.
The first scenario provides stronger synchrotron signals and stringent DM upper limits.
The irreducible level of magnetic turbulence set by the injection of $e^\pm$ from DM itself sets our conservative scenario, while we found that a magnetic field associated to stellar processes could only provide a small additional contribution that we disregard.
We excluded annihilation cross-sections above the thermal value for masses below 100 GeV in the optimistic magnetic field scenario. 
Our results surpass \textsc{Ret II} ATCA's limits \cite{regis2017dark}. 
The main results are reported in Figs.~\ref{fig:ULs_comp} and \ref{fig:ULs_decay}.\\
For imaging, Briggs (robust = 0) weighting provided the optimal balance between resolution and sensitivity. Both sensitivity and resolution proved to be critical: sensitivity sets the detection threshold, while resolution affects source subtraction accuracy and statistical power via independent beam count.  
As seen from Eq.~\ref{eq:sensitivity}, the sensitivity to DM signals can be improved with an increasing number of baselines, wider bandwidth, and longer integration time, which should be possible with the upcoming next-generation telescopes, like the SKA-Low and SKA-Mid telescopes. \\
Our results demonstrate that deep radio observations with MeerKAT achieve the most stringent DM radio limit from dSphs to date, paving the way for deep observations of the Local Group dwarf galaxies with the SKA.

\acknowledgments
The MeerKAT telescope is operated by the South African Radio Astronomy Observatory (SARAO), which is a facility of the South African National Research Foundation (NRF), an agency of the Department of Science and Innovation. This project made use of data from the open time project SCI-20220822-GB-01, where GB was the PI.\\
SS, GB, and SM acknowledge the support from the Pan African Planetary Science Network (PAPSSN) for a PhD at the University of the Witwatersrand, which enabled  this research.\\
SM acknowledges support from the NRF via SARAO through the Wits-SARAO Grant (UID: 97792). \\
MR acknowledges support by the European Union – Next Generation EU and by the Italian Ministry of University and Research (MUR) via the PRIN 2022 Project No. 20228WHTYC – CUP: D53C24003550006, and by the  Research grant TAsP (Theoretical Astroparticle Physics) funded by \textsc{infn}.\\

\providecommand{\mnras}{Mon. Not. Roy. Astron. Soc.}
\providecommand{\apj}{Astrophys. J.}
\providecommand{\apjs}{Astrophys. J. Suppl.}
\providecommand{\prd}{Phys. Rev. D}
\providecommand{\prl}{Phys. Rev. Lett.}
\providecommand{\aap}{Astron. Astrophys.}
\providecommand{\jcap}{JCAP}
\providecommand{\nat}{Nature}
\providecommand{\araa}{Ann. Rev. Astron. Astrophys.}
\providecommand{\pasp}{Publ. Astron. Soc. Pac.}

\bibliographystyle{JHEP}
\bibliography{Refs}
\end{document}